\documentclass[journal]{IEEEtran}

\usepackage{graphicx}  

\usepackage[centertags]{amsmath}   
\begin{document}
%
\title{
Traceable 2D finite-element simulation of the whispering-gallery modes of
axisymmetric electromagnetic resonators}
%
%
\author{Mark Oxborrow
\thanks{Manuscript revised Mar.~19, 2007.
This work was supported by the UK National Measurement
System's Quantum Metrology Programme. 
The author works at the National Physical Laboratory, Teddington, UK.}}
%
%
%
\markboth{Traceable 2D finite-element simulation of whispering-gallery modes ...}
{Oxborrow
: Traceable 2D finite-element simulation of WG modes ...}
%



\maketitle
\begin{abstract}
This paper explains how a popular, commercially-available software
package for solving partial-differential-equations (PDEs),
as based on the finite-element method (FEM), can be configured to
calculate, efficiently, the frequencies and fields of the whispering-gallery (WG) modes
of axisymmetric dielectric resonators. The approach is traceable;
it exploits the PDE-solver's ability to accept the definition
of solutions to Maxwell's equations in so-called `\emph{weak form}'.
Associated expressions and methods for estimating
a WG mode's volume, filling factor(s) and, in the case of closed(open)
resonators, its wall(radiation) loss, are provided.
As no transverse approximation is imposed, the approach remains accurate even for
\emph{quasi}-transverse-magnetic/electric modes of low, finite azimuthal mode order.
The approach's generality and utility are demonstrated by modeling several non-trivial structures:
(i)~two different optical microcavities  [one toroidal made of silica, the other
an AlGaAs microdisk]; (ii) a 3rd-order sapphire:air Bragg cavity;
(iii) two different cryogenic sapphire WG-mode resonators; both
(ii) and (iii) operate in the microwave X-band.
By fitting one of (iii) to a set of measured resonance frequencies, the
dielectric constants of sapphire at liquid-helium temperature have been
estimated.
\end{abstract}
%
\IEEEpeerreviewmaketitle
\section{Introduction\label{sec:Introduction}}
%
%
%
%
%
%
%

\PARstart{N}{on-trivial} electromagnetic structures can be modeled through
computer-aided design (CAD) tools in conjunction with
programs for numerically solving Maxwell's equations.
Though alternatives abound \cite{kunz93,boriskina03,ctyroky04},
the latter often use the finite-element method (FEM)
\cite{zienkiewicz00,taheri89}.
Within such a scheme, a problem frequently encountered when attempting
to determine the values of electromagnetic parameters from experimental data is
a lack of \emph{traceability}: significant dependencies between the data, the model's configurational
settings, and the inferred values of parameters cannot be adequately isolated,
understood, or quantified. Traceability demands that both the model's definition and its solution
remain amenable to complete, explicit description. And, furthermore, convenience requires that
the representations adopted for this purpose be concise --yet wholly unambiguous.
\subsection{Whispering-gallery-mode resonators\label{subsec:WhisperingGallery}}
\begin{figure}[h]
\centering
\includegraphics[width=0.68\columnwidth]{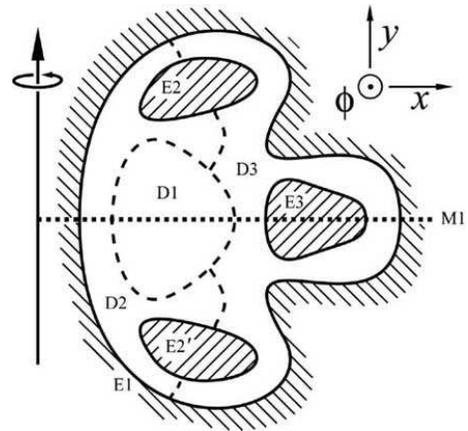}
\caption{Generic axisymmetric resonator in cross-section (medial half-plane).
A dielectric volume (in 3D) or `domain' (in 2D) is enclosed by an electric wall (E1)
--or one subject to some different boundary condition, as per subsubsection \ref{subsubsec:Open}.
This domain comprises several subdomains (D1, D2, and D3), each containing a spatially
uniform dielectric (that could be just free space). Some of these subdomains
(D2 and D3) are bounded internally by electric walls (E2, E2$'$ and E3).
The resonator has (optionally) a mirror symmetry through its (horizontal) equatorial
plane (dashed line M1); on imposing an electric or magnetic wall over this plane,
only either the upper or lower half of the resonator need be simulated.}
\label{fig:generic_resonator}
\end{figure}

Certain compact electromagnetic structures support closed whispering-gallery (WG) modes.
Though elliptical \cite{nockel97} or even non-planar (`crinkled' \cite{tobar01} or `spooled' \cite{sumetsky04})
WG morphologies exhibit advantageous features with respect to certain applications,
this paper considers only (closed, planar) WG modes with circular trajectories, as supported
by axisymmetric resonators within the class depicted in Fig.~\ref{fig:generic_resonator},
upon which various recent scientific innovations \cite{wolf04,rokhsari05,spillane05,srinivasan06}
are based.%
Many of the current commercial software packages for modeling electromagnetic
resonators suffer, however, from a `blind spot' when it comes to calculating,
efficiently (hence accurately), such resonators' whispering-gallery modes.
The popular MAFIA/CST package \cite{mafia} is a case in point:
as Basu \emph{et al} \cite{basu04} and no doubt others have experienced,
it cannot be configured to take advantage of a circular WG mode's known
azimuthal dependence, \emph{viz.}~$\textrm{exp}(\pm \textrm{i} M \phi)$, where $M$ (an
integer $\ge 0$) is the mode's azimuthal mode order, and $\phi$ the azimuthal coordinate.
Though frequencies and field patterns can be obtained (at least for WG modes
of low azimuthal mode order), the computationally advantageous reduction of the problem
from 3D to 2D that the resonator's rotational symmetry affords is, consequently,
precluded\footnote{
About the best one can do is simulate a `wedge' [over an azimuthal domain $\Delta  \phi = \pi/(2 M)$ wide]
between radial electric and magnetic walls.};
and the ability to simulate high-order WG modes with sufficient
accuracy (for metrological purposes) is, exasperatingly, lost.
\subsection{Brief, selected history of WG-mode simulation\label{subsec:BriefHistory}}
The method of `separating the variables' provides
analytical expressions for the WG modes of right-cylindrical
uniform dielectric cavities (or shells) \cite{wilson46,ramo84,inan00}.
By matching expressions across certain boundaries, approximate WG-mode solutions
for composite cylindrical cavities can be obtained \cite{tobar91a,tobar04,wolf04},
whose discrete/integer indices (related to symmetries)
provide a nomenclature \cite{krupka99a} for classifying the lower-order WG modes
of all similar structures.
Extensions of the basic mode-matching method encompassing spatially non-uniform field
polarizations have been developed \cite{ctyroky04}.

The accurate solution of arbitrarily shaped axisymmetrical
dielectric resonators requires numerical methods.
Apart from the finite-element method (FEM) itself,
the most developed and (thus) immediately exploitable alternatives include
(given here only for reference --not considered in any greater detail):
(i) the Ritz-Rayleigh variational or `moment' methods \cite{krupka94,krupka96,monsoriu02},
(ii) the finite difference time domain method (FDTD) \cite{kunz93,alford00}, and (iii)
the boundary-integral \cite{boriskina03} or boundary-element methods (BEM, including FEM-based
hybridizations thereof \cite{wolf03}). Zienkiewicz and Taylor \cite{zienkiewicz00},
particularly their table 3.2, indicate various commonalities between them (and FEM).\footnote{It
is remarked parenthetically here that
FDTD may be regarded as a variant of FEM employing local, discontinuous shape functions. It
is perhaps also worth acknowledging that, for resonators comprising just a few, large domains
of uniform dielectric, the boundary-integral methods (based on Green functions),
which --in a nutshell-- exploit such uniformity to reduce the problem's dimensionality by one,
will generally be more computationally efficient than FEM.}

The application of the finite-element method to the solving of Maxwell's equations
has a history \cite{rahman91}, and is now an industry \cite{mafia,COMSOL,ansys};
ref.~\cite{zienkiewicz00} supplies FEM's theoretical underpinnings.
Though the method can solve for all three of a WG mode's
field components, the statement of Maxwell's (coupled partial-differential) equations
in component form can be onerous, if not excluded outright by the equation-solving
software's lack of configurability.
With circular whispering-gallery modes, the configurational effort
can be significantly reduced by invoking a so-called
`transverse' approximation \cite{basu04,kippenberg04}, wherein only
a single (scalar) partial-differential equation is solved (in 2D).
Here, either the magnetic or electric field is assumed to
lie everywhere parallel to the resonator's axis of rotational symmetry
(see figure B.1 of ref.~\cite{kippenberg04}). This approximation is, however, uncontrolled.\footnote{It is noted paranthetically that
basic mode matching \cite{wolf04} also invokes the same transverse approximation and is thus equally uncontrolled.}
This paper demonstrates that, through only a modicum of extra configurational effort,
the transverse approximation and its associated doubts can be wholly obviated.

A problem that besets the direct application of
FEM to the solving of  Maxwell's equations is the generation of (many) spurious solutions \cite{auborg91,lee93},
associated with the local gauge invariance, or `null space' \cite{lee93}, that is a feature
of the equations' `curl' operators.
At least two research groups have nevertheless developed in-house software tools for
calculating the WG modes of axisymmetric dielectric resonators, that:
(i)~solve for all field components (\emph{i.e.}~no transverse approximation is invoked),
(ii)~are 2D (and thus numerically efficient) and (iii)
effectively suppress spurious solutions (without detrimental
side-effects) \cite{auborg91,krupka94,krupka96,osegueda94,santiago94}.
The method described in this paper sports these same three attributes.
With regard to (iii),~the approach adopted by Auborg \emph{et al}
was to use different finite elements (\emph{viz.} a mixture of  `Nedelec' and `Lagrange' -both 2nd order)
for different components of the electric and magnetic fields;
Osegueda \emph{et al} \cite{osegueda94}, on the other hand, used a so-called
`penalty term' to suppress (spurious) divergence of the magnetic field.
Stripping away its motivating remarks, applications and illustrations,
this paper, in essence, translates
the latter approach into explicit \emph{`weak-form'} expressions,
that can be directly and openly ported to any partial-differential equation (PDE) solver
(most notably COMSOL/FEMLAB \cite{COMSOL}) capable of accepting such.
\section{Method of solution\label{sec:Method}}
\subsection{Weak forms\label{subsec:WeakForms}}
\noindent \emph{Scope:}
The resonators treated below comprise volumes of dielectric space bounded
by either electric or magnetic walls (or both) --see again Fig.~\ref{fig:generic_resonator};
the restriction to axisymmetric resonators is only invoked at the start of subsection
\ref{subsec:Axisymmetric}.
The resonator's dielectric space comprises voids (\emph{i.e.}~free
space) and pieces of (sufficiently low-loss) dielectric material;
its (default) enclosing surfaces will generally be metallic, corresponding
to electric walls.
When modeling resonators whose forms exhibit reflection symmetries,
where the magnetic and electric fields of their solutions transform either symmetrically or
antisymmetrically through each mirror plane, perfect magnetic and electric walls can be
alternatively imposed over these planes to solve for different `sectors' of solutions.

The electromagnetic fields within the dielectric volumes of the resonator obey Maxwell's
equations \cite{bleaney76,inan00}, as they are applied to continuous, macroscopic media \cite{robinson73}.
Assuming the resonator's constituent dielectric elements have negligible (or at
least the same) magnetic susceptibility, the magnetic field strength $\textbf{H}$
will be continuous across interfaces.\footnote{The method described in this paper could be
to extended to treat resonators containing dielectrics with different magnetic susceptibilities by
setting up (within the PDE-solver --\emph{i.e.} COMSOL) `coupling variables' at interfaces.}
This property makes it advantageous to solve for $\textbf{H}$ (or, equivalently, the magnetic flux
density $\textbf{B} = \mu \textbf{H}$ --with a constant global magnetic permeability $\mu$),
as opposed to the electric field strength $\textbf{E}$ (or displacement $\textbf{D}$).
Upon substituting one of Maxwell's curl equations into the another,
the problem reduces to that of solving a (modified) vector Helmholtz equation:
\begin{equation}
\label{eq:helmholtz}
\boldsymbol{\nabla} \pmb{\times} ({\boldsymbol{\epsilon}^{-1}}\; \boldsymbol{\nabla} \pmb{\times} \textbf{H})
-\alpha \boldsymbol{\nabla} (\boldsymbol{\nabla} \cdot \textbf{H})
+ c^{-2} \partial^2\textbf{H}/{\partial t^2} = 0,
\end{equation}
subject to appropriate boundary conditions (read on). Here, $c$ is the speed of light
and ${\boldsymbol{\epsilon}^{-1}}$ the inverse relative permittivity tensor;
one assumes that the resonator's dielectric elements are linear, such that
${\boldsymbol{\epsilon}^{-1}}$ is a (tensorial) constant --\emph{i.e.}~independent of field strength.
Providing no magnetic monopoles lurk inside the resonator,
Maxwell's equations demand that $\textbf{H}$ be free of divergence,
\emph{i.e.}~$\boldsymbol{\nabla} \cdot \textbf{H} = 0$.
The middle, so-called `penalty' term on the left-hand side of
equation \ref{eq:helmholtz} acts to suppress spurious solutions,
for which (in general) $\boldsymbol{\nabla} \cdot \textbf{H} \ne 0$;
it has exactly the same form as that used by
Osegueda \emph{et al} \cite{osegueda94}.\footnote{Though COMSOL can implement
mixed (`Nedelec' plus `Lagrange') finite elements \cite{krupka94}, it was found
that equation~\ref{eq:helmholtz}'s penalty term (with its weighting factor
somewhere in the range $0.01 \leq \alpha \leq 10$) could, in conjunction with 2nd-order Lagrange
finite elements (applied to all three components of $\textbf{H}$), always satisfactorily
suppress the spurious modes.}
The constant $\alpha$ controls the penalty term's weight with respect to its
Maxwellian neighbors; $\alpha  = 1$ was taken for every simulation
presented in sections \ref{sec:ExampleApplications} and \ref{sec:PermDet}.

Reference~\cite{inan00} (particular section 1.3 thereof) supplies a primer
on the electromagnetic boundary conditions stated here.
The magnetic flux density at any point on a (perfect) electric wall
satisfies $\textbf{B} \cdot \textbf{n} = 0$,
where $\textbf{n}$ denotes the wall's surface normal vector.
Providing the magnetic susceptibility of the dielectric medium
bounded by the electric wall is not anisotropic, this condition is equivalent to
\begin{equation}
\textbf{H} \cdot \textbf{n} = 0.
\label{eq:electricwallH}
\end{equation}
The electric field strength at the electric wall obeys
\begin{equation}
\textbf{E} \pmb{\times} \textbf{n} = 0.
\label{eq:electricwallD}
\end{equation}
These two equations ensure that the magnetic(electric) field strength
is oriented tangential(normal) to the electric wall. As is pointed out
in reference~\cite{osegueda94}, equation \ref{eq:electricwallD} is a so-called
`natural' (or, synonymously, a `naturally satisfied') boundary condition
within the finite-element method --see ref.~\cite{zienkiewicz00}.

The boundary conditions corresponding to a perfect magnetic wall
(dual to the those for an electric wall) are
\begin{equation}
\label{eq:magneticwallD}
\textbf{D} \cdot \textbf{n} = 0,
\end{equation}
and
\begin{equation}
\label{eq:magneticwallH}
\textbf{H} \pmb{\times} \textbf{n} = 0;
\end{equation}
these two equations ensure that the electric displacement(magnetic field strength)
is oriented tangential(normal) to the magnetic wall. Again, the latter equation is naturally satisfied.

One now invokes Galerkin's method of weighted residuals \cite{zienkiewicz00};
ref.~\cite{lee93} provides an analogous treatment when solving for the
electric field strength ($\textbf{E}$). Both sides of equation \ref{eq:helmholtz} are multiplied
(scalar-product contraction) by the complex conjugate of a `test' magnetic field strength $\tilde{\textbf{H}}^*$,
then integrated over the resonator's dielectric volume. Upon expanding the permittivity-modified
`curl of a curl' operator (to extract a similarly modified Laplacian operator), then integrating by
parts (spatially), then disposing of surface terms through the electric- or magnetic-wall
boundary conditions stated above, one arrives (equivalent to equation (2) of reference \cite{osegueda94}) at
\begin{multline}
\label{eq:weakhelm}
\int_{\textrm{V}} [ (\boldsymbol{\nabla} \pmb{\times} \tilde{\textbf{H}}^*) \; \frac{\cdot}{\boldsymbol{\epsilon}} \;
(\boldsymbol{\nabla} \pmb{\times} \textbf{H}) -
~~~~~~~~~~\\
\; \alpha (\boldsymbol{\nabla} \cdot \tilde{\textbf{H}}^*) (\boldsymbol{\nabla} \cdot \textbf{H}) +
c^{-2} \tilde{\textbf{H}}^* \cdot \partial^2\textbf{H}/{\partial t^2} ] \, \textrm{d} \textrm{V} = 0,
\end{multline}
where `$\int_{\textrm{V}}$' denotes a volume integral over the resonator
and
$(\boldsymbol{\nabla} \pmb{\times} \tilde{\textbf{H}}^*) \frac{\cdot}{\boldsymbol{\epsilon}}
(\boldsymbol{\nabla} \pmb{\times} \textbf{H}) \equiv
\sum^3_{i, j = 1} [\boldsymbol{\nabla} \pmb{\times} \tilde{\textbf{H}}^*]_i \epsilon^{-1}_{ij}
[\boldsymbol{\nabla} \pmb{\times} \textbf{H}]_j$,
where
$\epsilon^{-1}_{ij}$ are the components of the inverse relative permittivity tensor.
The three terms appearing in the integrand correspond directly to the three weak-form terms required
to define an appropriate model within a partial-differential-equation solver.

Assuming that the physical dimensions and electromagnetic properties of the resonator's
components are temporally invariant (or at least `quasi-static'), solutions or `modes'
take the form $\textbf{H}(\textbf{r};t) = \textbf{H}(\textbf{r}) \textrm{exp}(-\textrm{i} 2 \pi f t)$,
where $\textbf{r}$ is the vector of spatial position,
$t$ the time, and $f$ the mode's resonance frequency.
The last, `temporal' term in equation \ref{eq:weakhelm}'s integrand can thereupon be
re-expressed as $-(\bar{c} f)^2 \tilde{\textbf{H}}(\textbf{r})^* \cdot \textbf{H}(\textbf{r})$,
where $\bar{c} \equiv 2 \pi/c$;
this re-expression reveals the integrand's complete dual symmetry
between $\tilde{\textbf{H}}^*$ and $\textbf{H}$.
\subsection{Axisymmetric resonators\label{subsec:Axisymmetric}}
The analysis is now restricted to axisymmetric resonators,
where a system of cylindrical coordinates (see top right Fig.~\ref{fig:generic_resonator}),
aligned with respect to the resonator's axis of rotational symmetry,
has components $\{r, \phi, z\}$
$\equiv \{$`rad(ial)', `azi(muthal)', `axi(ial)'$\}$.
The aim is to calculate the resonance frequencies and field patterns
of the resonator's circular WG modes, whose phase varies as
$\textrm{exp}(\textrm{i} M \phi)$, with $M =\{0, 1, 2, ...\}$
the mode's azimuthal mode order.\footnote{The method does not require $M$ to be large;
even modes that are themselves axisymmetric, corresponding to $M=0$,
such as the one shown in Fig.~\ref{fig:BraggCavity}(b), can be calculated.}
Viewed as a three-component vector field over (for the moment) a three-dimensional space,
the time-independent part of the magnetic field strength now takes the form
\begin{equation}
\label{eq:Hcylin}
\textbf{H}(\textbf{r}) =  \textrm{e}^{\textrm{i} M \phi} \left\{ \right.  H^r(r,z),
\textrm{i} \, H^\phi(r,z),
H^z(r,z) \left. \right\}
\end{equation}
where an `$\textrm{i}$' ($ \equiv \surd(-1)$) has been inserted into the field's azimuthal component
to allow, in subsequent solutions, all three component amplitudes
$\left\{ H^r, H^\phi, H^z \right\}$ to be each expressible as a real
amplitude multiplied by a common complex phase factor.
The relative permittivity tensor of an axisymmetric dielectric material is diagonal
with entries (running down the diagonal)
$\boldsymbol{\epsilon}_\textrm{diag.} = \{\epsilon_\perp, \epsilon_\perp, \epsilon_\parallel\}$,
where $\epsilon_\parallel$($\epsilon_\perp$) is the material's relative permittivity in the
axial direction (in the transverse or `perpendicular' plane --as spanned by its radial and azimuthal directions).

One now substitutes equation \ref{eq:Hcylin} into equation \ref{eq:weakhelm}'s integrand;
textbooks provide the required explicit expressions for
the vector differential operators in cylindrical coordinates \cite{ramo84,inan00}.
A radial factor, $r$, is included here from the volume
integral's measure: $\textrm{d} \textrm{V} = 2 \pi \, r \, \textrm{d} r \, \textrm{d} \phi \, \textrm{d} z$ (the
common factor of $2 \pi$ is dropped from all expressions below.)
The first, `Laplacian' weak term is given by
\begin{equation}
\label{eq:laplacianweakall}
(\boldsymbol{\nabla} \pmb{\times} \tilde{\textbf{H}}^*) \frac{\cdot}{\boldsymbol{\epsilon}} \;
(\boldsymbol{\nabla} \pmb{\times} \textbf{H}) =
\bigl( \frac{A}{r} + B + r \,C \bigr) / ({\epsilon_\perp \epsilon_\parallel}),
\end{equation}
where
\begin{multline}
\label{eq:laplacianweakA}
A \equiv \{\,\epsilon_\perp [\tilde{H}^\phi H^\phi
- M (\tilde{H}^\phi H^r + H^\phi \tilde{H}^r  ) + M^2 \tilde{H}^r H^r )] \\
~~~~~~+\epsilon_\parallel M^2 \tilde{H}^z H^z  \, \},
\end{multline}
\begin{multline}
\label{eq:laplacianweakB}
B \equiv \epsilon_\perp [ \tilde{H}^\phi_r ( H^\phi - M H^r ) + H^\phi_r (\tilde{H}^\phi - M \tilde{H}^r)]\\
~~~~~-\epsilon_\parallel M (\tilde{H}^z H^\phi_z  + H^z \tilde{H}^\phi_z  ),
\end{multline}
\begin{multline}
\label{eq:laplacianweakC}
C \equiv \{ \,\epsilon_\perp \tilde{H}^\phi_r H^\phi_r \\
~~+\epsilon_\parallel [(\tilde{H}^z_r - \tilde{H}^r_z) (H^z_r-H^r_z) + \tilde{H}^\phi_z H^\phi_z ] \, \}.
\end{multline}
Notation: $H^\phi_r$ denotes the partial derivative of $H^\phi$ (the azimuthal component
of the magnetic field strength) with respect to $r$ (the radial component of displacement),
\emph{etc.}. Here, the individual factors and terms have been ordered and grouped
so as to display the dual symmetry. Similarly, the weak penalty term is given by
\begin{equation}
\label{eq:penaltyweakall}
\alpha (\boldsymbol{\nabla} \cdot \tilde{\textbf{H}}^*) (\boldsymbol{\nabla} \cdot \textbf{H})
= \alpha \{ \frac{D}{r} + E + r \,F \},
\end{equation}
where
\begin{multline}
\label{eq:penaltyweakD}
D \equiv \tilde{H}^r H^r  - M ( \tilde{H}^\phi H^r + H^\phi\tilde{H}^r )+ M^2 \tilde{H}^\phi H^\phi,
\end{multline}
\begin{multline}
\label{eq:penaltyweakE}
E \equiv  (\tilde{H}^r_r + \tilde{H}^z_z) (H^r - M H^\phi)\\
+ (\tilde{H}^r - M \tilde{H}^\phi) (H^r_r + H^z_z), ~~~~~~~~~~~
\end{multline}
\begin{equation}
\label{eq:penaltyweakF}
F \equiv ( \tilde{H}^r_r + \tilde{H}^z_z) (H^r_r + H^z_z).~~~~~~~~~~~~~~~~~~~~~~~~~~~
\end{equation}
And the temporal weak-form (`dweak') term is given by
\begin{multline}
\label{eq:temporalweak}
\tilde{\textbf{H}}^* \cdot \partial^2\textbf{H}/{\partial^2 t} =
c^{-2} \, r \, ( \tilde{H}^r  H^r_{tt}  + \tilde{H}^\phi H^\phi_{tt} + \tilde{H}^z H^z_{tt}) ~~~~~~~~~~~~~~~~~~~\\
~~~~~~~~~ = -\bar{c}^2 f^2 \, r \, ( \tilde{H}^r  H^r + \tilde{H}^\phi H^\phi + \tilde{H}^z H^z),
\end{multline}
where $H^r_{tt}$ denotes the double partial derivative of $H^r$ w.r.t.~time, \emph{etc.}.
Note that none of the terms on the right-hand sides of equations
\ref{eq:laplacianweakall} through \ref{eq:temporalweak} depend on the azimuthal coordinate $\phi$;
the problem has been reduced from 3D to 2D.
\subsection{Axisymmetric boundary conditions\label{subsec:AxisymmetricBCs}}
An axisymmetric boundary's unit normal in cylindrical components can
be expressed as $\{n_r, 0, n_z\}$ --note vanishing azimuthal component.
The electric-wall boundary conditions, in cylindrical components, are as follows:
$\textbf{H} \cdot \textbf{n} = 0$ gives
\begin{equation}
\label{eq:electricwallHcylcomp}
H^r n_r + H^z n_z = 0,
\end{equation}
and $\textbf{E} \pmb{\times} \textbf{n} = 0$ gives both
\begin{equation}
\label{eq:electricwallEcylcomp1}
H^r_z - H^z_r = 0
\end{equation}
and
\begin{equation}
\label{eq:electricwallEcylcomp2}
\epsilon_\perp (H^\phi -H^r M + H^\phi_r r) n_r - \epsilon_\parallel(H^z M - H^\phi_z r) n_z = 0.
\end{equation}
When the dielectric permittivity of the medium bounded by the electric wall is isotropic,
the last condition reduces to
\begin{equation}
\label{eq:electricwallEcylcomp2isotrop}
(H^\phi -H^r M + H^\phi_r r) n_r - (H^z M - H^\phi_z r) n_z = 0.
\end{equation}
%
%
%
%
%
%
%
The magnetic-wall boundary conditions, in cylindrical components,
are as follows: $\textbf{D} \cdot \textbf{n} = 0$ gives
\begin{equation}
\label{eq:magneticwallDcylcomp}
(H^z M - H^\phi_z r) n_r + (H^\phi -H^r M + H^\phi_r r) n_z = 0,
\end{equation}
and $\textbf{H} \pmb{\times} \textbf{n} = 0$ gives both
\begin{equation}
\label{eq:magneticwallHcylcomp1}
H^z n_r -H^r n_z = 0
\end{equation}
and
\begin{equation}
\label{eq:magneticwallHcylcomp2}
H^\phi =0.
\end{equation}
Note that the transformation
$\{ n_r \rightarrow - n_z, n_z \rightarrow n_r\}$,
connects equations \ref{eq:electricwallHcylcomp} and \ref{eq:magneticwallHcylcomp1},
and equations \ref{eq:electricwallEcylcomp2isotrop} with \ref{eq:magneticwallDcylcomp}.
The above weak-form expressions and boundary conditions, \emph{viz.}~equations \ref{eq:laplacianweakall}
through \ref{eq:magneticwallHcylcomp2} are the key
results of this paper: once inserted into a PDE-solver, the WG modes of axisymmetric dielectric
resonators can be readily calculated.
\section{Postprocessing of solutions\label{sec:Postprocessing}}
Having determined, for each mode, its frequency and all three components of
its magnetic field strength $\textbf{H}$ as a function of position,
other relevant fields and parameters can be derived from this information.
\subsection{Other fields (related through Maxwell's equations)\label{subsec:Maxwellian}}
Straightaway, the magnetic flux density $\textbf{B} = \mu \textbf{H}$.
As no real (`\emph{non}-displacement') current flows within a dielectric,
$\nabla \pmb{\times} \textbf{H}(t) = \partial \textbf{D}(t) / \partial t $,
thus $\textbf{D} = -\textrm{i} (2 \pi f)^{-1} \nabla \pmb{\times} \textbf{H}(t)$.
And, $\textbf{E} = \boldsymbol{\epsilon}^{-1}\textbf{D}$, where
$\boldsymbol{\epsilon}^{-1}$ is the (diagonal) inverse permittivity tensor,
as already discussed in connection with equation \ref{eq:weakhelm} above.
\subsection{Mode volume\label{sec:ModeVolume}}
Accepting various caveats (most fundamentally, the problem of mode-volume divergence --see footnote 8)
as addressed by Kippenberg \cite{kippenberg04}, the volume of a mode is defined as \cite{srinivasan06}
\begin{equation}
\label{eq:mode_volume}
V_{\rm{mode}} = \frac{\int \int \int_{\rm{b.-s.}} \epsilon |\textbf{E}|^2 \textrm{d} \textrm{V} }
{ {\rm{max}} [\epsilon |\textbf{E}|^2 ]},
\end{equation}
where ${\rm{max}} [ ... ]$, denotes the maximum value of its functional argument,
and $\int \int \int_{\rm{b.-s.}} ... \textrm{d} \textrm{V} $ denotes integration over and around
the mode's `bright spot' --where its electromagnetic field energy is concentrated.
\subsection{Filling factor\label{sec:Fillingfactor}}
The resonator's electric filling factor, for a given dielectric component (labeled `$\rm{diel.}$'),
a given mode, and a given field direction, (`$\rm{dir.}$' $ \in \{\rm{radial}, \rm{azimuthal}, \rm{axial}\}$),
is defined as
\begin{equation}
\label{eq:filling_factor}
F_{\rm{diel.}}^{\rm{dir.}} = \frac{\int \int \int_{\rm{diel.}} \epsilon_{\rm{pol.}} |E^{\rm{dir.}}|^2 \textrm{d} \textrm{V} }
{\int \int \int \epsilon |\textbf{E}|^2 \textrm{d} \textrm{V} },
\end{equation}
where $\int \int \int_{\rm{diel.}} ... \textrm{d} \textrm{V} $ denotes integration
over the component's volume and
$\rm{pol.}$ = $\{\perp,\parallel\}$ for $\rm{dir.}$ = $\{\rm{radial} \; \rm{or} \; \rm{azimuthal},$ $\rm{axial}\}$.
The numerators and denominators of equations \ref{eq:mode_volume} and \ref{eq:filling_factor}
can be readily evaluated using the PDE-solver's post-processing features.
\subsection{Wall loss (closed resonators)\label{subsec:WallLosses}}
Real resonators suffer losses that render the $Q$s of their modes finite.
The energy stored in a mode's electromagnetic field is
$U = (1/2) \int \int \int \mu |\textbf{H}|^2 \textrm{d} \textrm{V}$.
For axisymmetric resonators, the 3D volume integral
$\int \int \int \textrm{d} \textrm{V}$ reduces to a 2D integral
$\int \int (2 \pi r) \textrm{d} r \textrm{d} z$ over the resonator's
medial half-plane.
The surface current induced in the resonator's enclosing
electric wall is (see ref.~\cite{inan00}, page 205, for example)
$\textbf{J}_s = \textbf{n} \pmb{\times} \textbf{H}$;
the averaged-over-a-cycle power dissipated in the wall is
$P_{\rm{loss}} = (1/2) \int \int  R_s |\textbf{H}_{\rm{t}}|^2 \textrm{d} \textrm{S}$,
where $\textbf{H}_{\rm{t}}$ is the tangential (with respect to the wall)
component of $\textbf{H}$,
$R_s = (\pi f \mu / \sigma)^{1/2}$ is the wall's surface resistivity
(see refs.~\cite{inan00,bleaney76}), $\sigma$ the wall's (bulk) electrically conductivity,\footnote{It is
here pointed out that, at liquid-helium temperatures, the bulk and surface resistances of metals
can depend greatly on the levels of (magnetic) impurities within them \cite{pobell92}, and the
text-book $f^{-1/2}$ dependence of surface resistance on frequency is often
violated \cite{fletcher94}.}
and $f$ the mode's frequency.
For axisymmetric resonators, the 2D surface integral $\int \int \textrm{d} \textrm{S}$
reduces to a 1D integral $\int (2 \pi r) \textrm{d} \textrm{l}$ around the periphery
of the resonator's medial (r-z) half-plane.
The quality factor, defined as $2 \pi f \, U / P_{\rm{loss}}$,
due to the wall loss can thus be expressed as
\begin{equation}
\label{eq:Q_wall_loss}
Q_{\rm{wall}}= \frac{2 \pi f \, \mu}{R_{\rm{s}}} \Lambda = (4 \pi f \, \mu \, \sigma)^{1/2} \Lambda,
\end{equation}
where $\Lambda$, which has the dimensions of a length, is defined as
\begin{multline}
\label{eq:length}
\Lambda = \frac{\int \int \int |\textbf{H}|^2 \textrm{d} \textrm{V} }
{\int \int |\textbf{H}_{\rm{t}}|^2 \textrm{d} \textrm{S} }\\
= \frac{\int \int [(H^r)^2 + (H^\phi)^2 + (H^z)^2] \, r \, \textrm{d} r \,\textrm{d} z}
{\int [|H^\phi|^2 + |H^z n_r - H^r n_z|^2] r\, \rm{dl}}.
\end{multline}
Again, both integrals (numerator and denominator), hence  $Q_{\rm{wall}}$ itself, can be readily evaluated
through the PDE-solver's post-processing features.
\subsection{Radiation loss (open resonators)\label{subsec:RadLoss}}
\emph{Preliminary remarks:}
With \emph{open} whispering-gallery-mode resonators (either microwave \cite{bourgeois04b}
or optical \cite{rokhsari05,srinivasan06}), the otherwise highly localized WG mode
spreads throughout free-space;\footnote{As understood by Kippenberg \cite{kippenberg04},
this observation implies that the
support of equation \ref{eq:mode_volume}'s $\int \int \int_{\rm{b.-s.}} ... \textrm{d} \textrm{V} $ integral
(spanning the WG mode's bright spot) must be somehow limited, spatially, or otherwise (asymptotically) rolled
off, lest the integral diverge. [The so-called `quantization volume' associated
with the radiation extends to infinity.]} energy flows
away from the mode's bright spot (where the electric- and magnetic-field
amplitudes are greatest) as \emph{radiation}.
The tangential electric and magnetic fields on any closed surface surrounding
the bright spot constitute, by the `Field Equivalence Principle' \cite{schelkunoff36,balanis97}
(as a formalization of Huygen's picture), a secondary source of this radiation.
\subsubsection{Underestimator of loss via retro-reflection\label{subsubsec:retro-refl}}
Consider a closed, equivalent of the open resonator, with an enclosing
electric wall in the WG-mode's radiation zone. The wall's form is chosen such that
--as far as possible-- the open resonator's radiation propagates (as a predominantly
transverse and locally plane wave) in a direction that is locally normal to the wall.
The electric wall then reflects the otherwise open resonator's radiation
back onto itself --so creating a \emph{standing wave}, \emph{i.e.} a loss-less mode.
Through an argument reminiscent of Schelkunoff's induction theorem \cite{schelkunoff36,schelkunoff39},
the tangential magnetic field of this mode, $\textbf{H}_{\rm{t}}^{\rm{closed}}$, at any point just inside
the closed resonator's electric wall, can be related to that of the corresponding open resonator's
radiation, $\textbf{H}_{\rm{t}}^{\rm{open}}$, at the same point,
through $\textbf{H}_{\rm{t}}^{\rm{open}} > 2 \textbf{H}_{\rm{t}}^{\rm{closed}}$.
The radiation loss can be evaluated by integrating the cycle-averaged
Poynting vector over the electric wall,
\emph{i.e.} $P_{\rm{rad.}} = (1/2) \int \int  z_0 |\textbf{H}_{\rm{t}}^{\rm{open}}|^2 \textrm{d} \textrm{S} $,
where $z_0$ is the impedance of free space.
A bound on the mode's radiative $Q$-factor can thus be expressed as
\begin{equation}
Q_{\rm{rad.}} > (8 \pi f / c) \Lambda,
\label{eq:Q_rad_electric_wall}
\end{equation}
approaching equality when the WG mode's bright spot lies (in effect)
at an antinode of the mode's standing wave; $\Lambda$ here is exactly
that given by equation \ref{eq:length},
with $\textbf{H} = \textbf{H}_{\rm{t}}^{\rm{closed}}$, only now
the enclosing electric wall is in the radiation zone.
\subsubsection{Overestimator of loss via outward-going free-space impedance matching\label{subsubsec:Open}}
A complementary bound can be constructed by replacing the above closed resonator's
electric wall with one, of the same form, that attempts to match, impedance-wise,
the open resonator's radiation --and thus absorb it.
For transverse, locally plane-wave radiation in the radiation zone (in free space)
sufficiently far from the resonator,
the required impedance-matching boundary condition on the wall is
$z_0\, \textbf{n} \pmb{\times} \textbf{H} = \textbf{E} - \textbf{n} (\textbf{E} \cdot \textbf{n})$,\footnote{Note that
the direction (polarization) of $\textbf{E}$ or $\textbf{H}$ in the wall's plane is not constrained;
the two fields need only be orthogonal with their relative amplitudes in the ratio of the
impedance of free space $z_0$.}
where $\textbf{n}$ is the wall's inward-pointing normal.
Upon differentiating with respect to time and using Maxwell's displacement-current equation,
this condition can,
for a given mode, be generalized to
\begin{multline}
\label{eq:rad_match_mix}
\cos(\theta_{\rm{mix}}) \{\nabla \pmb{\times} \textbf{H} - \textbf{n} [(\nabla \pmb{\times} \textbf{H}) \cdot  \textbf{n}]\}\\
+ \sin(\theta_{\rm{mix}}) \, \textrm{i} \, \bar{c} f_{\rm{mode}} \, \textbf{n} \pmb{\times} \textbf{H} = 0,
\end{multline}
where $f_{\rm{mode}}$ is the mode's frequency,
$\bar{c} \equiv 2 \pi / c$ as before, and $\theta_{\rm{mix}}$ is a `mixing angle';\footnote{The
two terms on the left-hand side of equation
\ref{eq:rad_match_mix} can be viewed as implementing electric- (cf.~equation \ref{eq:electricwallD})
and magnetic-wall (cf.~equation \ref{eq:magneticwallH}) boundary conditions, respectively.
The (composite) boundary condition can be continuously adjusted between these two cases by varying
the mixing angle $\theta_{\rm{mix}}$. [Parenthetically:
$\theta_{\rm{mix}} = -\pi/4$ corresponds to impedance matching inward-coming
(as opposed to an outward-going) radiation.]}
for impedance matching (with respect to an outward-going radiation),
$\theta_{\rm{mix}} = \pi/4$.
Unless $\theta_{\rm{mix}} = N \pi/2$ for integer $N$, the $\textrm{i} = \sqrt{-1}$
in equation \ref{eq:rad_match_mix} breaks the hermitian-ness of the matrix that
the PDE-solver is required to eigensolve, leading to decaying modes with complex
eigenfrequencies $f_{\rm{mode}}$, and corresponding quality factors equal
\cite{srinivasan06} to
$\Re[f_{\rm{mode}}]/2\Im[f_{\rm{mode}}]$, where $\Re[...]$ and
$\Im[...]$ denote real and imaginary parts.
Without fine tuning, the enclosing wall's shape will not everywhere
lie exactly orthogonal to the direction of propagation of the WG mode's radiation;
thus, even for $\theta_{\rm{mix}} = \pi/4$, the radiation will not be completely
absorbed at the wall. A bound on the mode's radiative $Q$-factor can thus
be expressed as
\begin{equation}
\label{eq:Q_rad_match}
Q_{\rm{rad.}} < \Re[f_{\rm{mode}}]/2\Im[f_{\rm{mode}}],
\end{equation}
approaching equality on perfect absorption (no reflections).
\section{Example applications\label{sec:ExampleApplications}}
The source codes and configuration scripts used to implement the simulations
presented in this and the next section are freely available from the author.
\subsection{`Sloping-shoulders' cryogenic sapphire microwave resonator [UWA]\label{subsec:SlopingShoulders}}
\begin{figure}[h]
\centering
\begin{tabular}{@{}r@{\quad}l}
a:$\:$\mbox{\includegraphics[width=0.40\columnwidth]{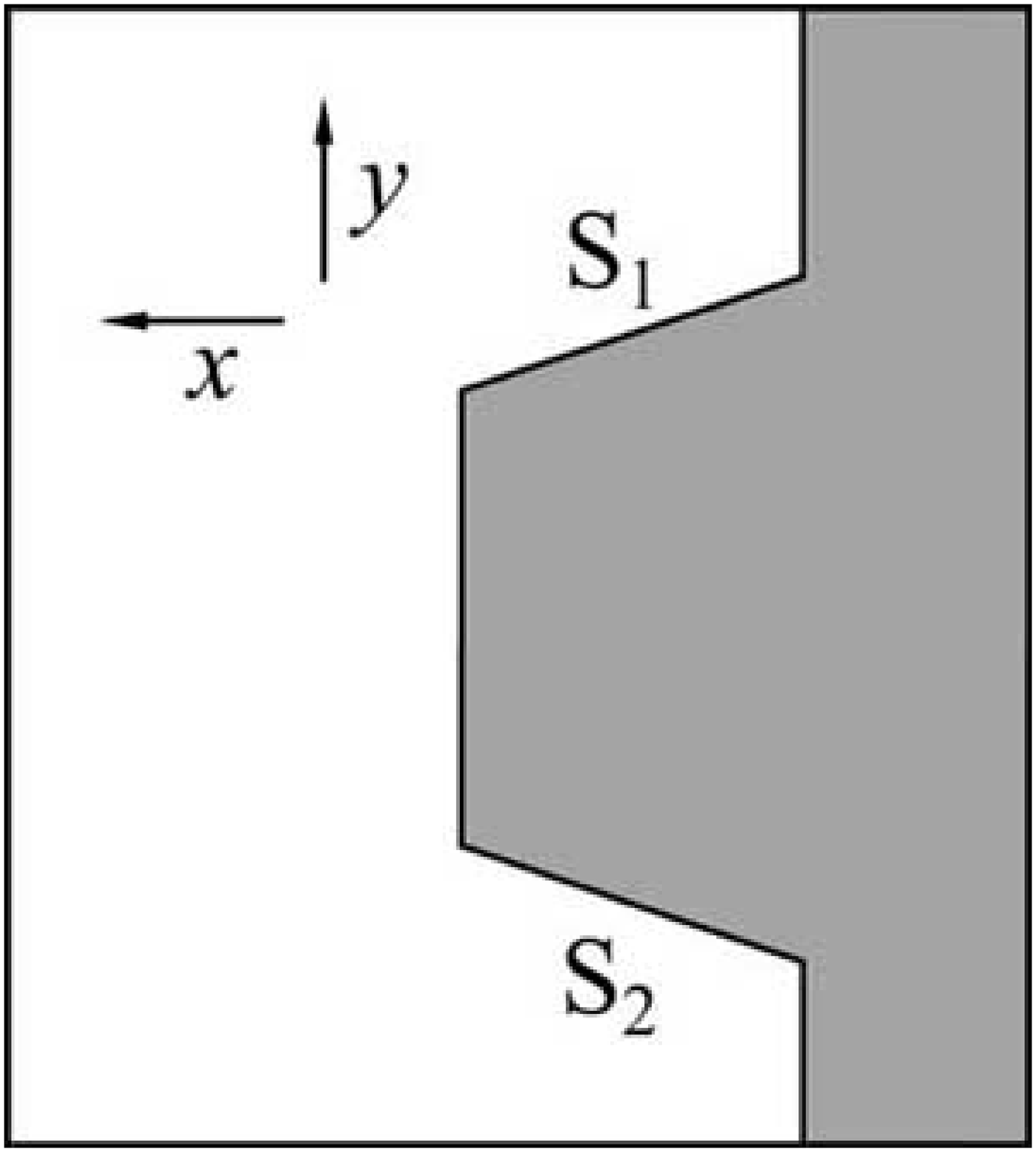}}&
b:$\:$\mbox{\includegraphics[width=0.40\columnwidth]{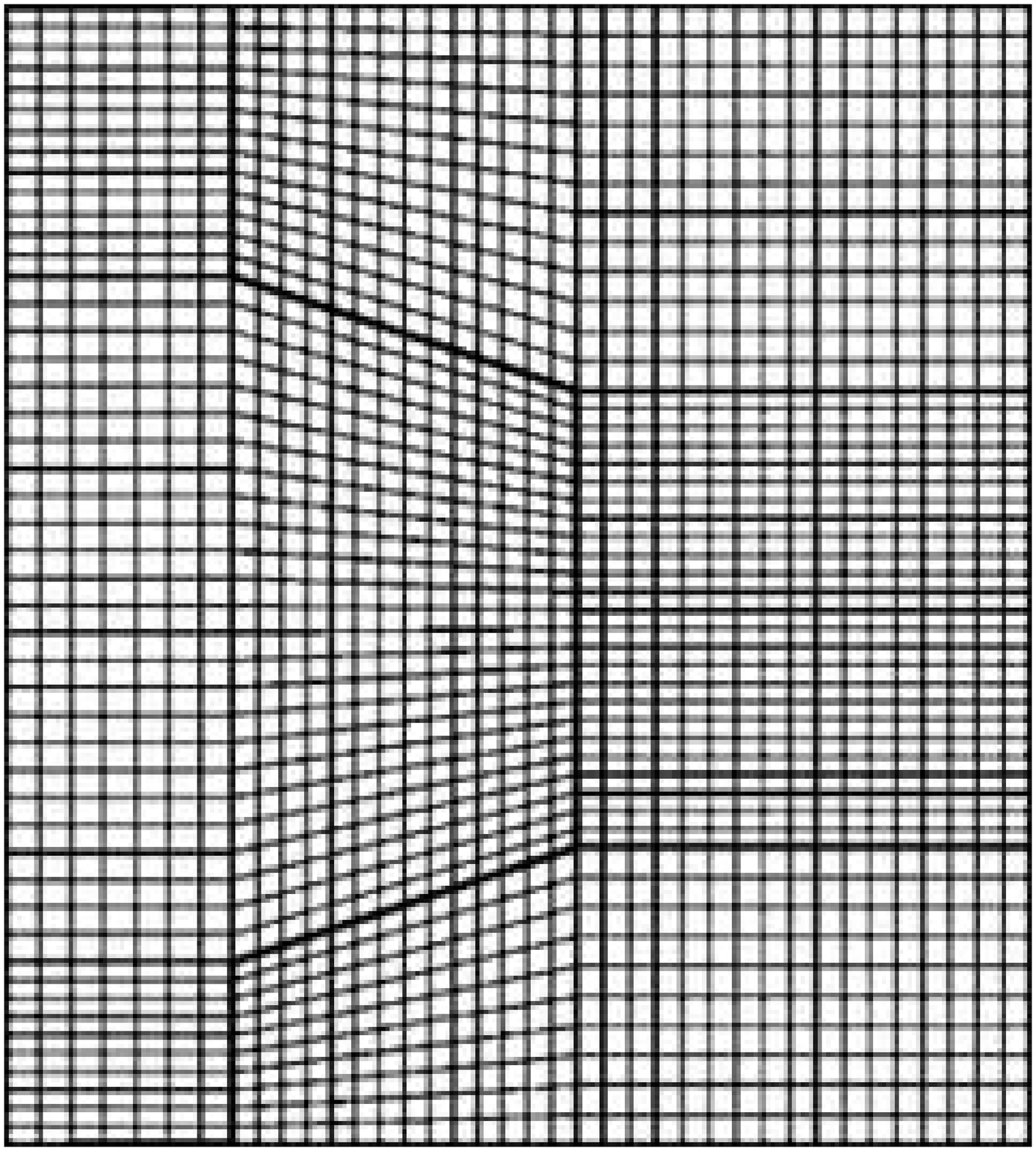}}\\
& \\
c:$\:$\mbox{\includegraphics[width=0.40\columnwidth]{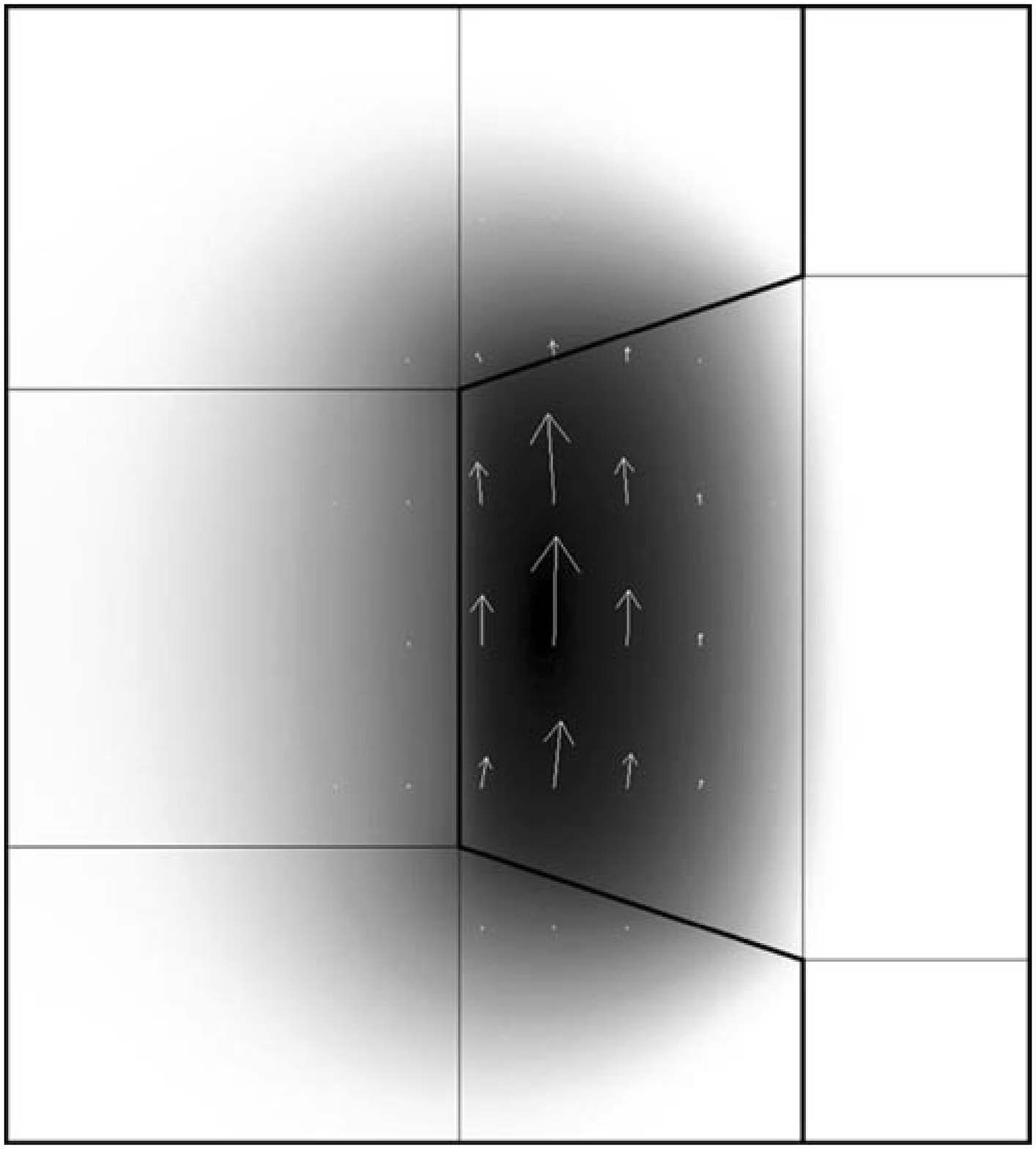}}&
d:$\:$\mbox{\includegraphics[width=0.40\columnwidth]{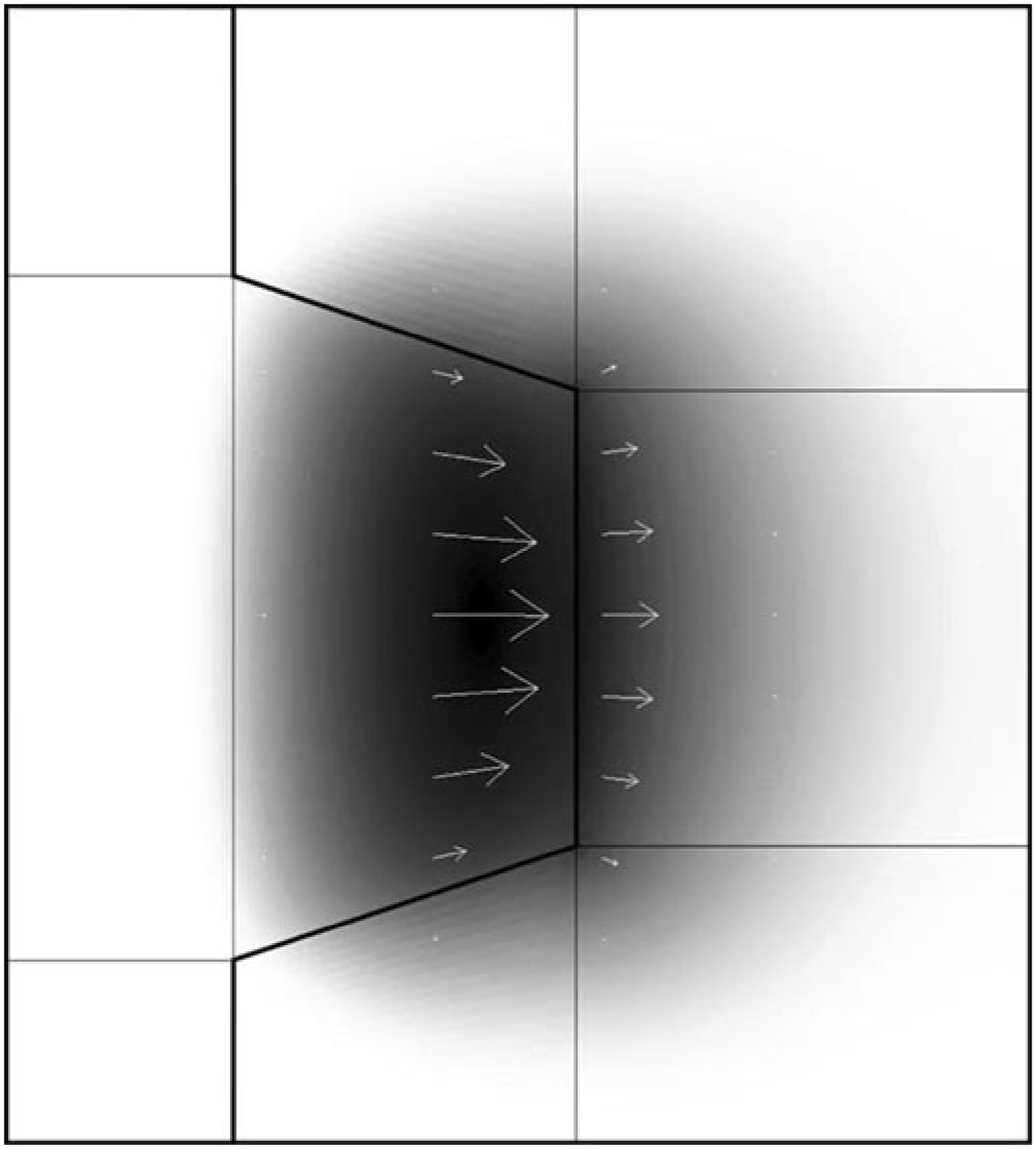}}\\
\end{tabular}
\caption{UWA's sloping-shoulder cryogenic sapphire resonator: (a) medial cross-section
through its geometry; the grey(white) shading corresponds to sapphire(vacuum); \textrm{S}$_1$ and
\textrm{S}$_2$ indicate the sapphire piece's upper and lower `shoulders'.
(b) mesh of the resonator's model structure generated by the FEM-based PDE-solver;
for clarity, only every other meshing line is drawn [\emph{i.e.}~(b) displays
the `half-mesh']; within (c) and (d), the (logarithmic) grey scale reflects the absolute
value of the vectorial magnetic $\textbf{H}$ and electric $\textbf{E}$ fields, respectively;
white arrows indicate the magnitude and direction of each field's medial component.}
\label{fig:UWAres}
\end{figure}
This axisymmetric resonator \cite{luiten95}
comprises a piece of monocrystalline sapphire mounted (co-axially)
within a cylindrical metal can --see Fig.~\ref{fig:UWAres}(a);
the crystal's optical (or `c') axis lies parallel to the resonator's geometric axis.
The piece's sloping shoulders (S$_1$ and S$_2$ \emph{ibid.}) make accurate simulation
via mode-matching less straightforward.
The resonator's form, as encoded into the PDE-solver, is taken from figure 3 of
ref.~\cite{wolf04}\footnote{It is remarked here that the drawn shape of the sapphire piece in
figure 3 of ref.~\cite{wolf04} is not wholly consistent with its given dimensions: its outer axial
sidewall is too long and the slope of its shoulders too slight.},
with the piece's outer diameter, the length of its outer axial sidewall,
the axial extent of each sloping shoulder, and the radius of each of its two spindles
equal to, at liquid-helium temperature (\emph{i.e.}~including cryogenic
shrinkages --see section \ref{sec:PermDet}) 49.97, 19.986, 4.996, and 19.988~mm, respectively.
The sapphire crystal's cryogenic permittivities were taken to be $\{\epsilon_\perp ,
\epsilon_\parallel \} = \{9.2725, 11.3486\}$, as given in ref.~\cite{krupka99a}.
Since the sapphire piece and its surrounding metal can do not share a common transverse
(`equatorial') mirror plane, the speeding up of the simulation through the imposition
of a magnetic or electric wall on such a plane (so
halving the 2D region to be analyzed) is precluded.

Fig.~\ref{fig:UWAres}(b) displays the FEM-based PDE-solver's meshing of the model resonator's structure;
in COMSOL's vernacular\footnote{The size/complexity of a finite-element mesh is quantified, within COMSOL
Multiphysics, by (i) the number of elements that go to compose its so-called `base mesh' and (ii) its total
number of degrees of freedom (DOF) --as associated with its so-called `extended mesh'.},
the mesh comprises 7296 base-mesh elements and 88587 degrees of freedom (`DOF').
It took typically 75~seconds, to obtain the resonator's lowest (in frequency) 16 modes,
for a single, given azimuthal mode order $M$, at [with respect to Fig.~\ref{fig:UWAres}(b)] full mesh
density, on a standard, 2004-vintage personal computer (2.4~GHz, Intel Xeo CPU), without altering the PDE-solver's
default eigensolver settings. With the azimuthal mode order set at $M = 14$, the model resonator's WGE$_{\rm{14,0,0}}$ mode was
found to lie at 11.925 GHz, to be compared with 11.932~GHz found experimentally \cite{wolf04}.

\noindent\emph{Wall loss:}
This mode's characteristic length $\Lambda$, as evaluated with respect to the can wall's enclosing surface,
was determined to be $2.6 \times 10^4$~m.
Based on ref.~\cite{fletcher94}, one estimates the surface resistance of copper at liquid-helium temperature
to be $7 \times 10^{-3} \; \Omega$ per square at 11.9 GHz, leading to a
wall-loss $Q$ of  $3.5 \times 10^{11}$ for the  WGE$_{\rm{14,0,0}}$ mode.

\noindent \emph{Filling factor:} Using equation \ref{eq:filling_factor}, the electric filling factors for the WGE$_{14,0,0}$ mode
were evaluated. These factors, presented in TABLE~\ref{tab:UWAfillfac} above, are in good agreement with those
labeled `FE' in Table~IV of ref.~\cite{wolf04}, that were obtained via a wholly independent FEM simulation of the same resonator.
\begin{table}[t]
\centering
\caption{Electric filling factors for the WGE$_{\rm{14,0,0}}$ mode of UWA's sloping-shoulders resonator }
\begin{tabular}{@{}r|lll}
$F_{\rm{diel.}}^{\rm{dir.}}$        & radial        & azimuthal                 & axial \\
\hline
sapphire                            & 0.80922       & 0.16494                   & $7.016 \times 10^{-3}$ \\
vacuum                              & 0.01061       & $8.0533 \times 10^{-3}$   & $1.6543 \times 10^{-4}$ \\
\end{tabular}
\label{tab:UWAfillfac}
\end{table}
\subsection{Toroidal silica optical resonator [Caltech]\label{subsec:SilicaToroidal}}
\begin{figure}[h]
\centering
\begin{tabular}{@{}l}
a:$\:$ \mbox{\includegraphics[width=0.70\columnwidth]{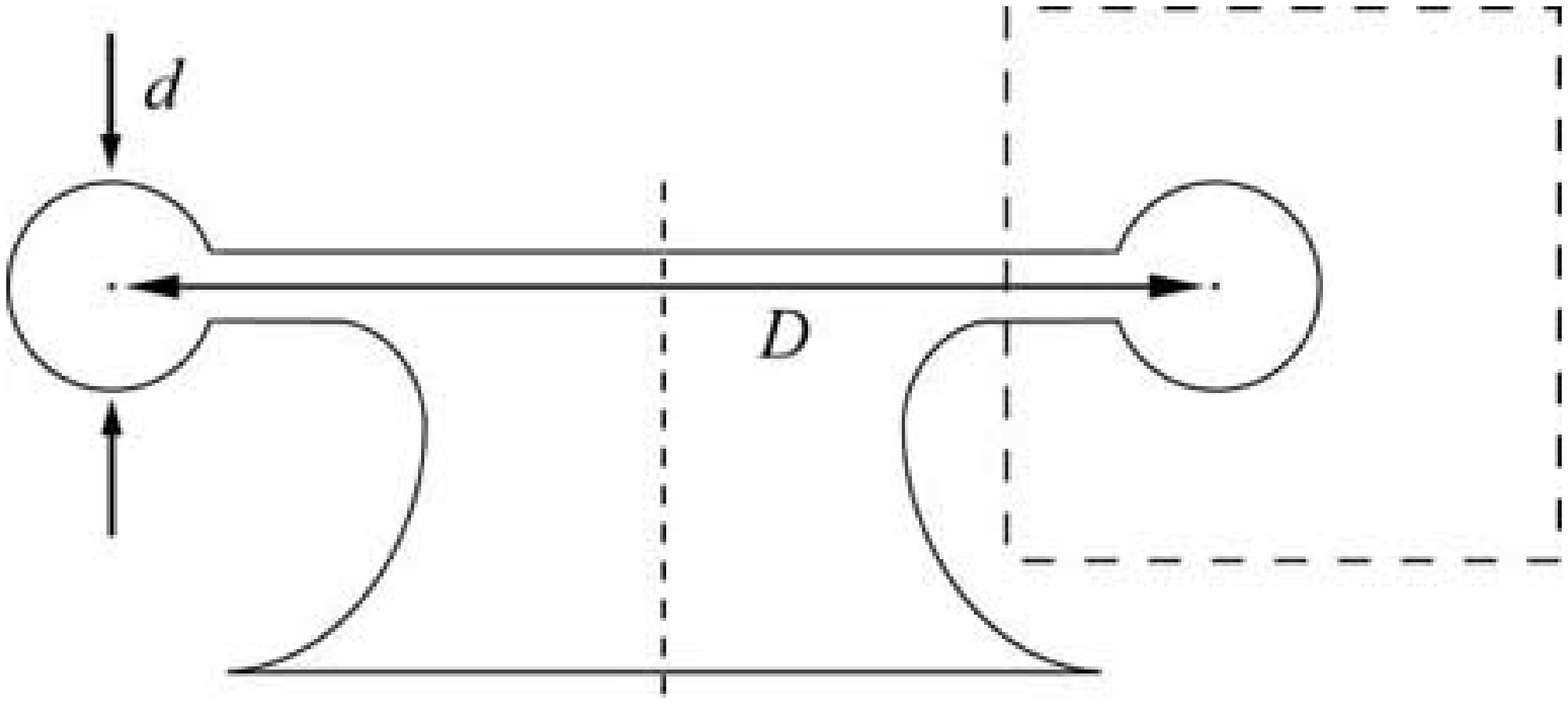}}\\
\\
b:$\:$ \mbox{\includegraphics[width=0.50\columnwidth]{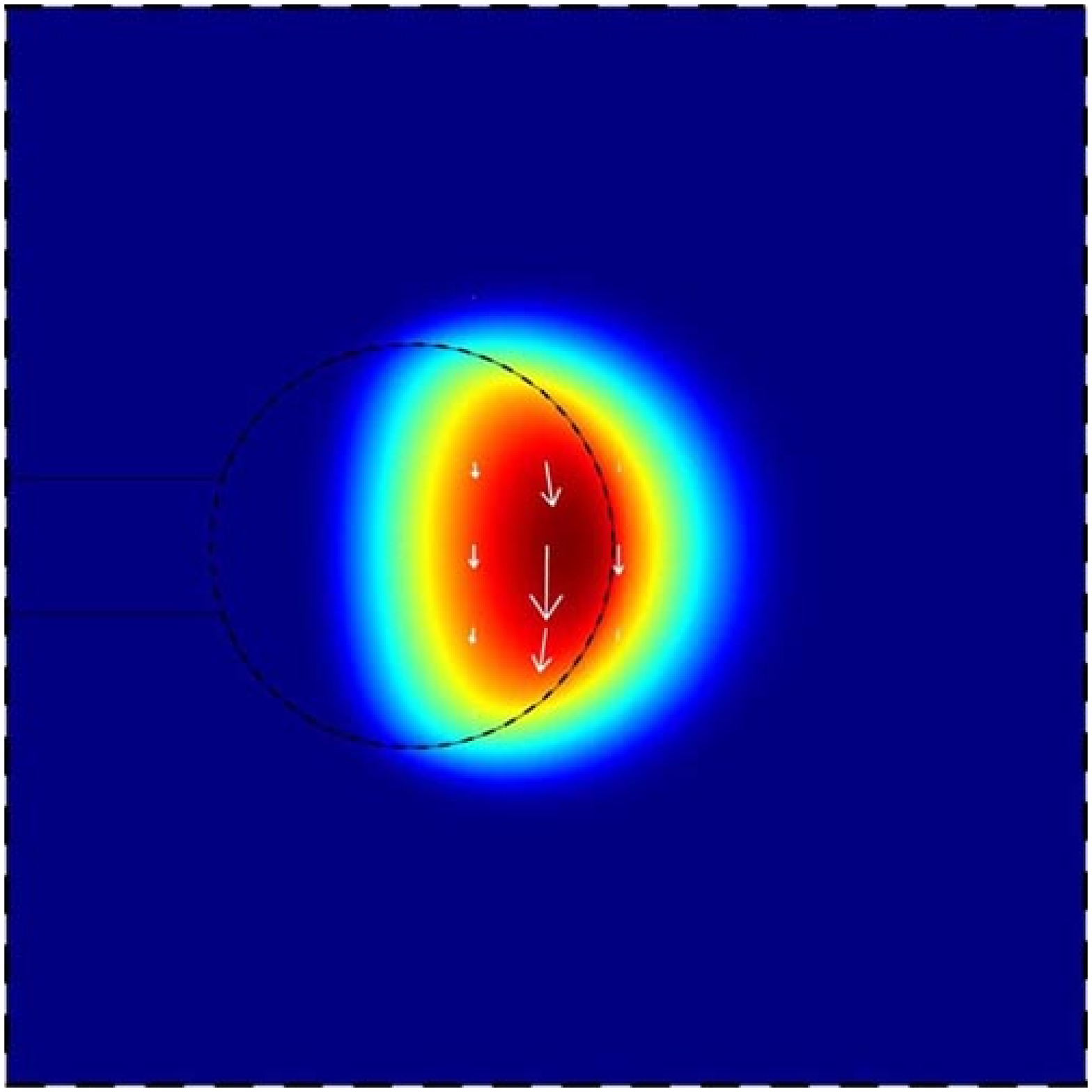}}\\
\end{tabular}
\caption{(a) Geometry (medial cross-section) and dimensions of a model toroidal silica microcavity resonator
--after ref. \cite{spillane05}; the torus' principal diameter $D = 16$ $\mu$m and its
minor diameter $d = 3$ $\mu$m; the central vertical dashed line indicates the resonator's axis of continuous
rotational symmetry.
(b) False-color surface plot of the (logarithmic) electric-field intensity $|\textbf{E}|^2$ over the dashed
box appearing in (a) for this resonator's TE$_{p=1,m=93}$ whispering-gallery mode; white arrows indicate
the electric field's magnitude and direction in the medial plane.}
\label{fig:ToroidalSilica}
\end{figure}
The resonator modeled here, based on ref.~\cite{spillane05}, comprises a silica
toroid, supported above a substrate by an integral interior `web';
its geometry is shown in Fig.~\ref{fig:ToroidalSilica}(a).
The toroid's principal and minor diameters
are $\{D,d\}$ = $\{16,3\}$ $\mu$m, respectively. The silica dielectric is presumed
to be wholly isotropic (\emph{i.e.}, no significant residual stress) with a relative permittivity
of $\epsilon_{\rm{sil.}} = 2.090$, corresponding to a refractive index of
$n_{\rm{sil.}} = \sqrt{\epsilon_{\rm{sil.}}} = 1.4457$ at the resonator's
operating wavelength (around 852~nm) and temperature.
The FEM model's pseudo-random triangular mesh
covered an 8-by-8 $\mu$m square [shown in dashed outlined
on the right of Fig.~\ref{fig:ToroidalSilica}(a)],
with an enhanced meshing density over the silica circle and
its immediately surrounding free-space; in total,
the mesh comprised 5990 (base-mesh) elements, with DOF~$= 36279$.
Temporarily adopting Spillane \emph{et al}'s terminology, the resonator's fundamental TE-polarized
93rd-azimuthal-mode-order mode (where `TE' here implies that the polarization of the
mode's electric field is predominantly aligned with the toroid's principal axis --\emph{not} transverse to it)
was found to have a frequency of $3.532667 \times 10^{14}$~Hz, corresponding to a
free-space wavelength of $\lambda$ = 848.629 nm (thus close to 852 nm).
Using equation \ref{eq:mode_volume}, this mode's volume was evaluated
to be 34.587 $\mu$m$^3$; if, instead, the definition stated in equation 5 of
ref.~\cite{spillane05} is used, the volume becomes
72.288 $\mu$m$^3$ --\emph{i.e.}~a factor of $n_{\rm{sil.}}^2$ greater.
These two values straddle the volume
of $\sim$55~$\mu$m$^3$, for the same dimensions of silica toroid
and (TE) mode-polarization, as inferred by eye from
figure 4 of ref.~\cite{spillane05}.
\subsection{Conical microdisk optical resonator [Caltech]\label{subsec:GaAlAsMicrodisk}}
The mode volume can be reduced by going to smaller resonators, which, unless
the optical wavelength is commensurately reduced, implies lower azimuthal mode orders.
\begin{figure}[h]
\centering
\begin{tabular}{@{}l}
a:$\:$ \mbox{\includegraphics[width=0.78\columnwidth]{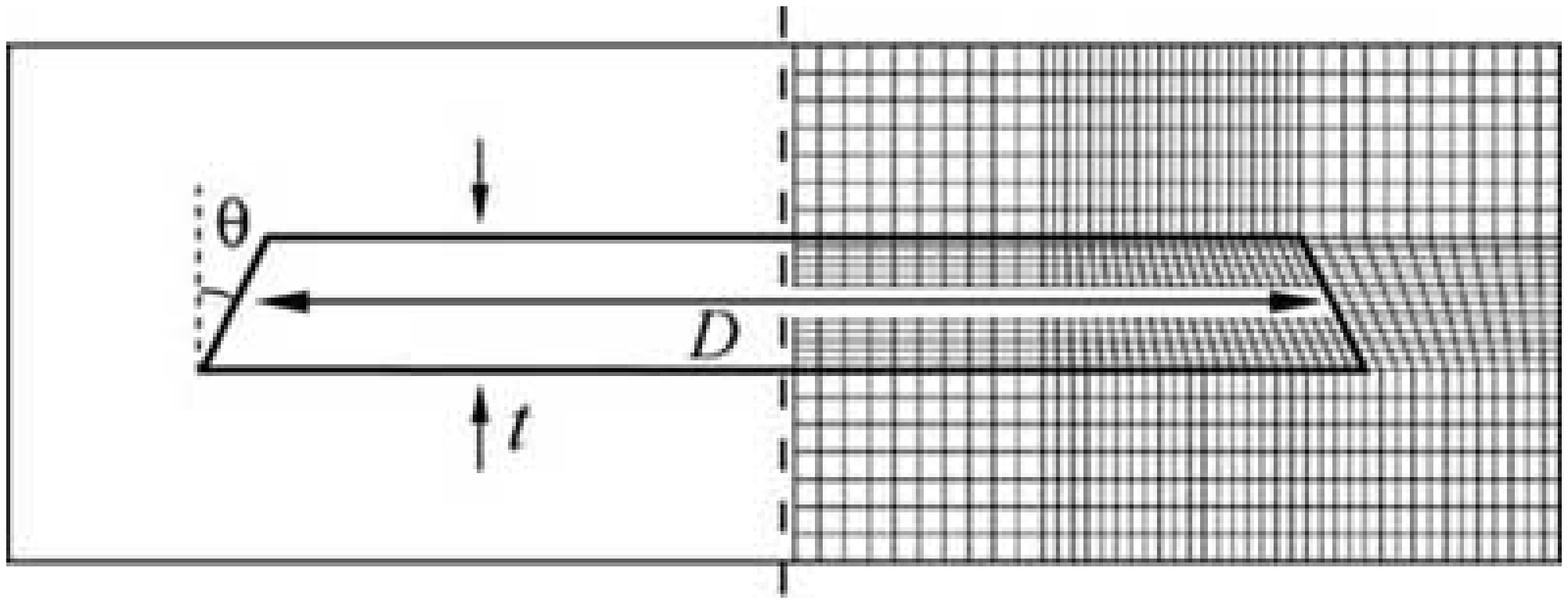}}\\
\\
b:$\:$ \mbox{\includegraphics[width=0.65\columnwidth]{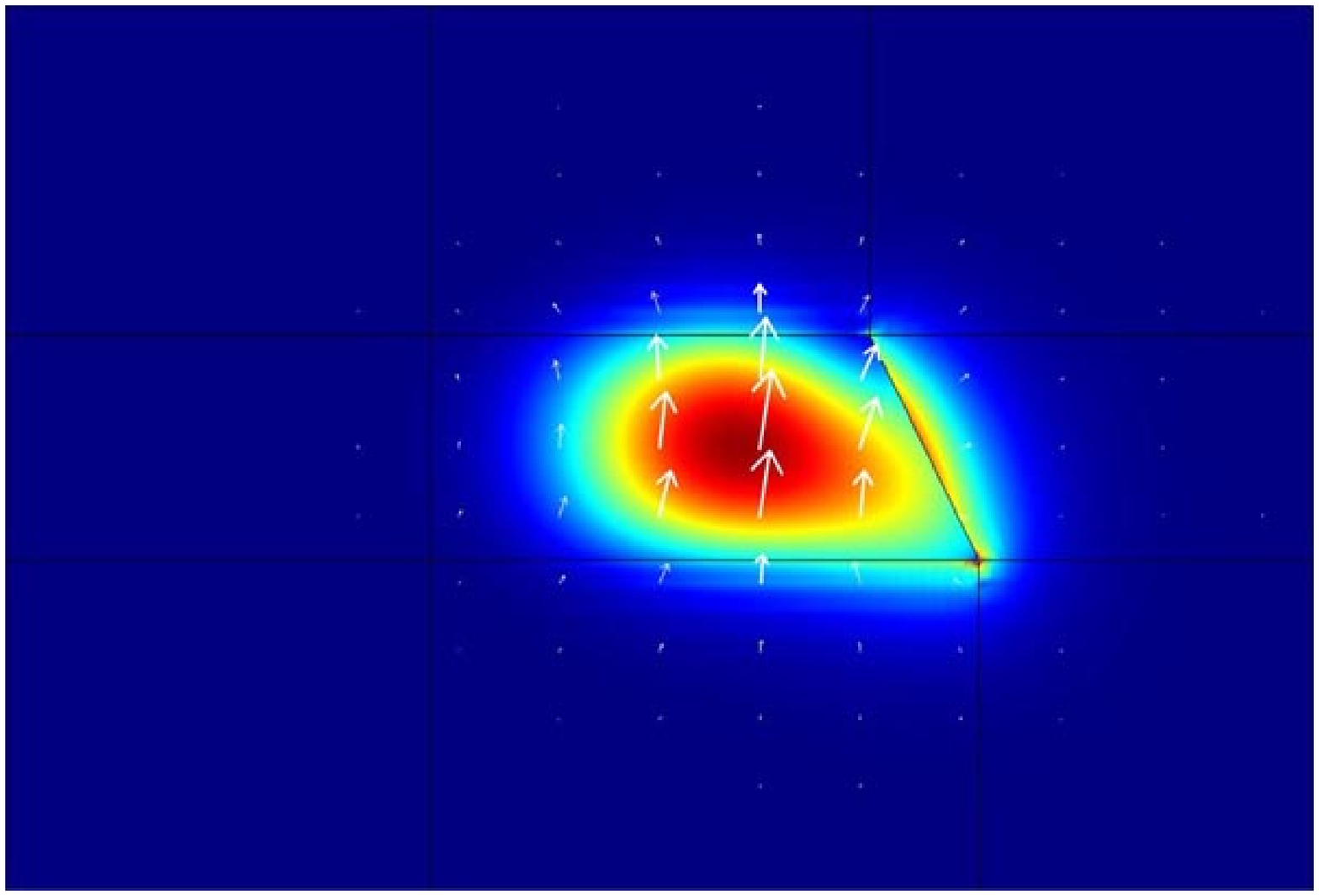}}\\
\end{tabular}
\caption{(a) Geometry (medial cross-section) and (half-)meshing of model GaAlAs microdisk resonator
--after ref. \cite{srinivasan06}; the disk's median diameter is $D$ = 2.12 $\mu$m and its
thickness (axial height) $t$ = 255 nm; its conical external sidewall subtends an angle
$\theta$ = 26$^\circ$ to the disk's (vertical) axis; for clarity, only every other line of the true
(full) mesh is drawn. The modeled domain in the medial half-plane is a rectangle stretching from
0.02 to 1.5 $\mu$m in the radial direction and from -0.5 to +0.5 $\mu$m in the axial direction.
(b) False-color surface plot of the (logarithmic) electric-field intensity
$|\textbf{E}|^2$ for the resonator's TE$_{p=1,m=11}$ mode at $\lambda$ = 1263.6 nm; again,
white arrows indicate the electric field's magnitude and direction in the medial plane.}
\label{fig:GaAlAsMicrodisk}
\end{figure}
The model `microdisk' resonator analyzed here, as depicted in Fig.~\ref{fig:GaAlAsMicrodisk}(a),
duplicates the structure defined in figure 1(a) of Srinivasan \emph{et al}~\cite{srinivasan06};
the disk's constituent dielectric (alternate layers of GaAs and GaAlAs) is
approximated as a spatially uniform, isotropic dielectric, with a refractive index equal to $n$ = 3.36.
The FEM-modeled domain comprised 4928 quadrilateral
base-mesh elements, with DOF~$=60003$. Adopting the same authors'
notation, the resonator's TE$_{p=1,m=11}$ whispering-gallery mode,
as shown in Fig.~\ref{fig:GaAlAsMicrodisk}(b), was found
at $2.372517 \times 10^{14}$ Hz, equating to $\lambda$ = 1263.6~nm;
for comparison, Srinivasan \emph{et al} found an equivalent mode at 1265.41~nm
[as depicted in their figure 1(b)].
It is pointed out here that the white electric-field arrows in Fig.~\ref{fig:GaAlAsMicrodisk}(b) [and also, though to a lesser extent,
in Fig.~\ref{fig:ToroidalSilica}(b)] are \emph{not} perfectly vertical --as the transverse approximation taken in references
\cite{wolf04,kippenberg04} would assume them to be; the \emph{quasi}-ness of the true mode's transverse-electric polarization
is thus revealed.

\noindent \emph{Mode volume:}
Using equation \ref{eq:mode_volume}, but with the mode excited as a
standing-wave (doubling the numerator while quadrupling the denominator),
the mode volume is determined to be $0.1484 \times \mu$m$^3 \simeq 2.79 (\lambda/n)^3$,
still in good agreement with Srinivasan \emph{et al}'s $\sim \! 2.8 (\lambda /n)^3$.

\begin{figure}[h]
\centering
\begin{tabular}{@{}l@{\quad}l@{\quad}l}
a:$\:$ \mbox{\includegraphics[width=0.25\columnwidth]{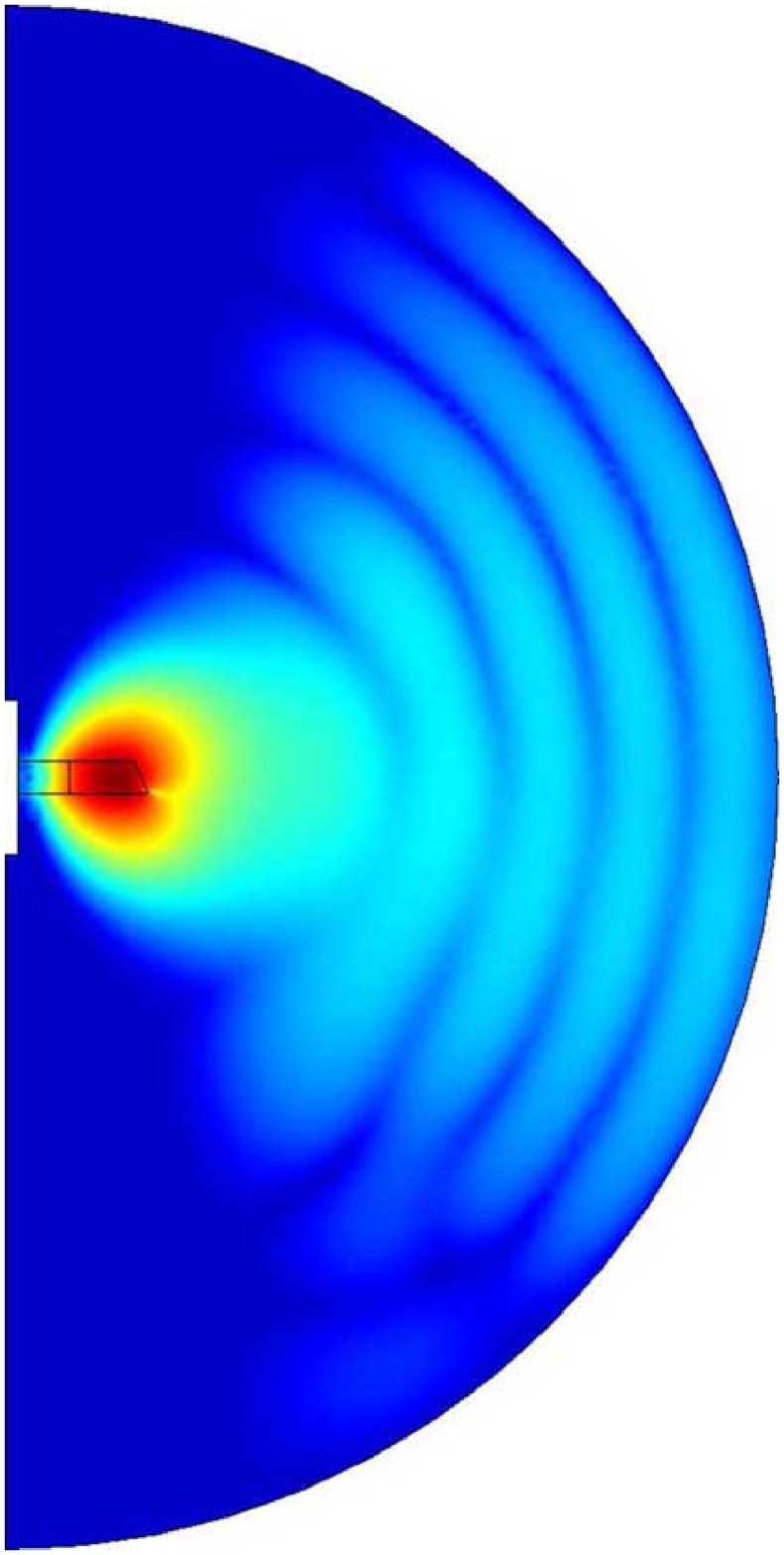}}
& b:$\:$ \mbox{\includegraphics[width=0.25\columnwidth]{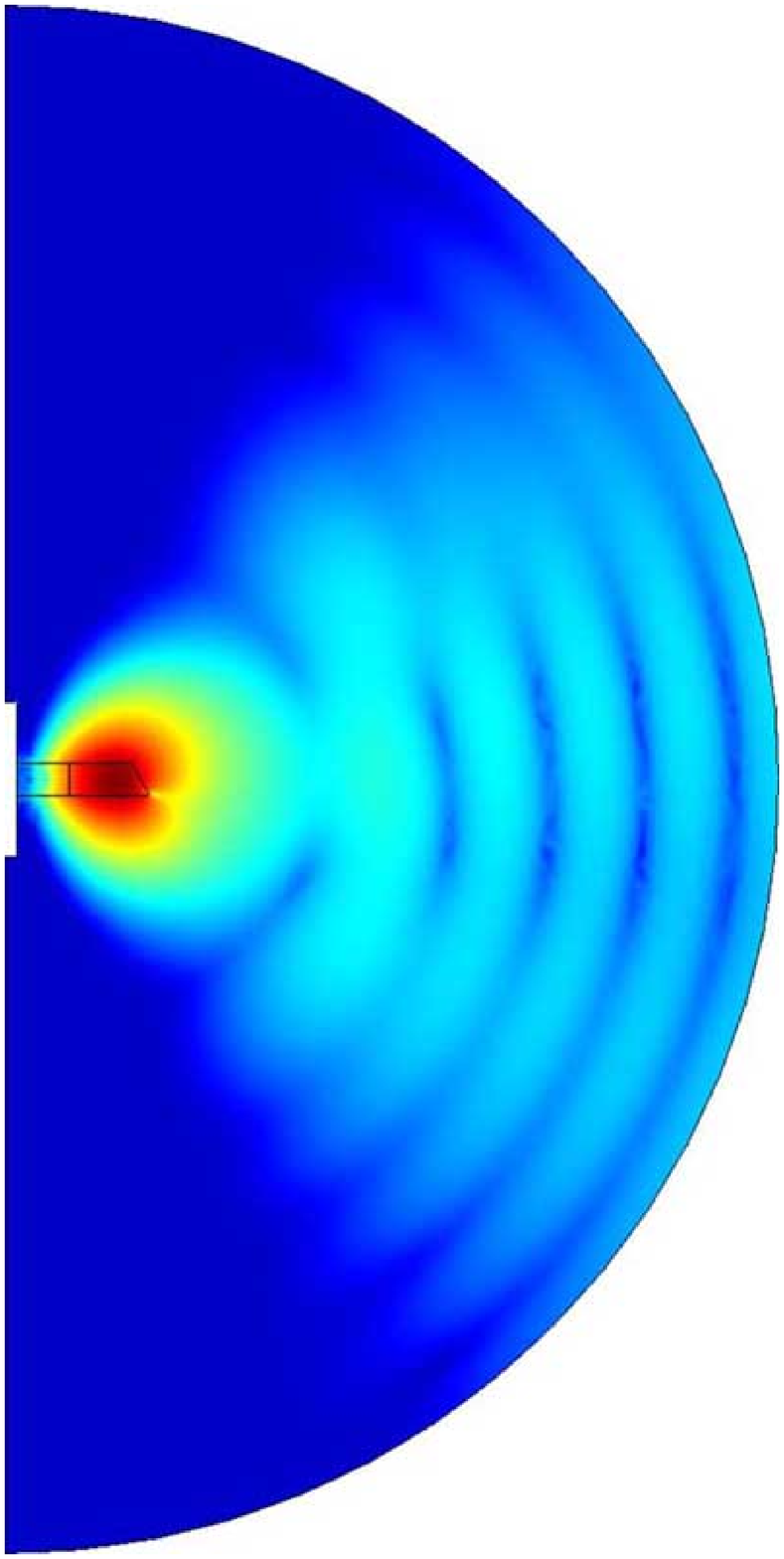}}
& c:$\:$ \mbox{\includegraphics[width=0.25\columnwidth]{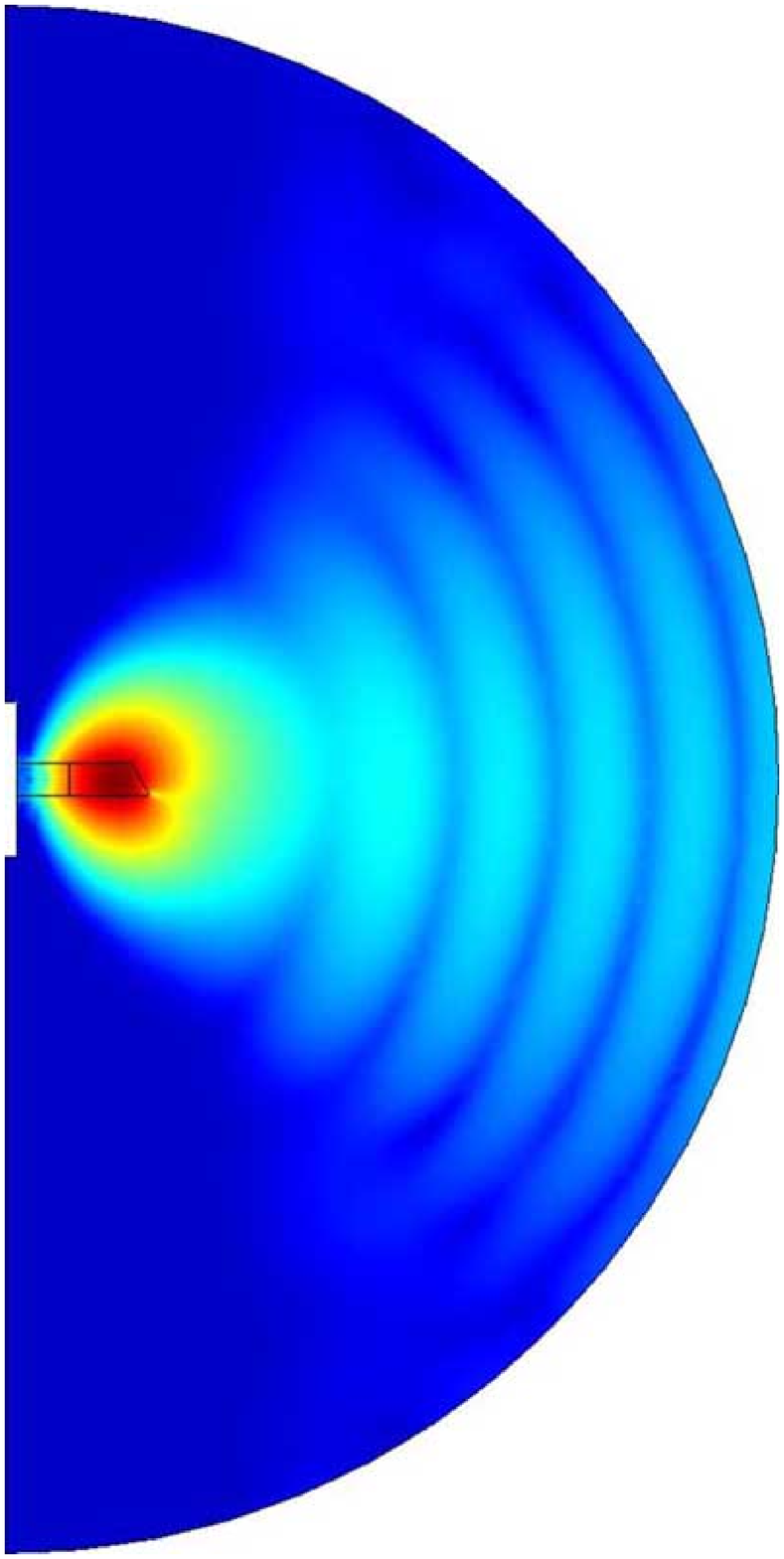}}\\
\end{tabular}
\caption{Radiation associated with the same [TE$_{p=1,m=11}$, $\lambda$ = 1263.6 nm] whispering-gallery mode
as presented in Fig.~\ref{fig:GaAlAsMicrodisk} (all three maps use the same absolute false-color scale):
(a) standing-wave (equal outward- and inward-going) radiation with the outer semicircular boundary set as a
magnetic wall; (b) the same but now with the boundary set as an electric wall; (c)
somewhat traveling (more outward- than inward-going) radiation with the boundary's impedance
set to that of an outward-going plane-wave in free space (and with the normal magnetic field
constrained to vanish). That (c)'s radiation field is somewhat dimmer than (b)'s is consistent
with the two different estimates of the resonator's radiative $Q$ corresponding to (b) and (c) [see text].}
\label{fig:RadiationLoss}
\end{figure}

\noindent \emph{Radiation loss:}
The TE$_{p=1,m=11}$ mode's radiation loss was estimated by implementing both the upper- and
lower-bounding estimators described in subsection \ref{subsec:RadLoss}. Here,
the microdisk and its mode were modeled over an approximate sphere, equating to
a half-disk in 2D (medial plane). The half-disk's diameter was 12~$\mu$m and
different electromagnetic conditions were imposed on its
semicircular boundary--see Fig.~\ref{fig:RadiationLoss}.\footnote{It is acknowledged that,
in reality, the microdisk's substrate would occupy a considerable part of half-disk's lower quadrant.}
With an electric-wall condition (\emph{i.e.} equations \ref{eq:electricwallH}
and \ref{eq:electricwallD} or, equivalently, \ref{eq:electricwallHcylcomp}
and \{\ref{eq:electricwallEcylcomp1}, \ref{eq:electricwallEcylcomp2isotrop}\})
imposed on the half-disk's entire boundary [as per Fig.~\ref{fig:RadiationLoss}(b)],
the right-hand of equation \ref{eq:Q_rad_electric_wall} was evaluated. And,
with the $\textbf{E} \pmb{\times} \textbf{n} = 0$ condition (\emph{viz.} equation \ref{eq:electricwallD})
on the boundary's semicircle replaced by the outward-going-plane-wave(-in-free-space)
impedance-matching condition (\emph{viz.}~equation \ref{eq:rad_match_mix} with $\theta_{\rm{mix}} = \pi/4$),
while the $\textbf{H} \cdot \textbf{n} = 0$ condition (equation \ref{eq:electricwallH})
is maintained, the right-hand side of equation \ref{eq:Q_rad_match} was evaluated
for the radiation pattern displayed in Fig.~\ref{fig:RadiationLoss}(c).
For a pseudo-random triangulation mesh comprising 4104 elements, with DOF~$ = 24927$,
the PDE solver took, on the author's office computer, 6.55 and 13.05 seconds, corresponding
to Figs.~\ref{fig:RadiationLoss}(b) and (c), respectively\footnote{The complex arithmetic
associated with the impedance-matching boundary condition meant that the PDE solver's
eigen-solution took approximately twice as long to run with this condition imposed --as compared to
pure electric- or magnetic-wall boundary conditions that do not involve complex arithmetic.},
to calculate 10 eigenmodes around
$2.373 \times 10^{14}$ Hz, of which the TE$_{p=1,m=11}$ mode was one.
Together, the resultant estimate on the TE$_{p=1,m=11}$ mode's radiative-loss quality factor is
$(1.31 < Q_{\rm{rad.}} < 3.82) \times 10^7$,
to be compared with the estimate of
$9.8 \times 10^6$ (at 1265 nm) reported in table 1 of ref.~\cite{srinivasan06}.
\subsection{Distributed-Bragg-reflector microwave resonator\label{subsec:BraggCav}}
The method's ability to simulate axisymmetric resonators of arbitrary cross-sectional complexity
is demonstrated here by simulating the 10-GHz TE$_{01}$ mode of a distributed-Bragg-reflector (DBR)
resonator as analyzed by Flory and Taber (F\&T) \cite{flory97} through mode matching.
\begin{figure}[h]
\centering
\begin{tabular}{@{}r@{\quad}l}
a:$\:$\mbox{\includegraphics[width=0.45\columnwidth]{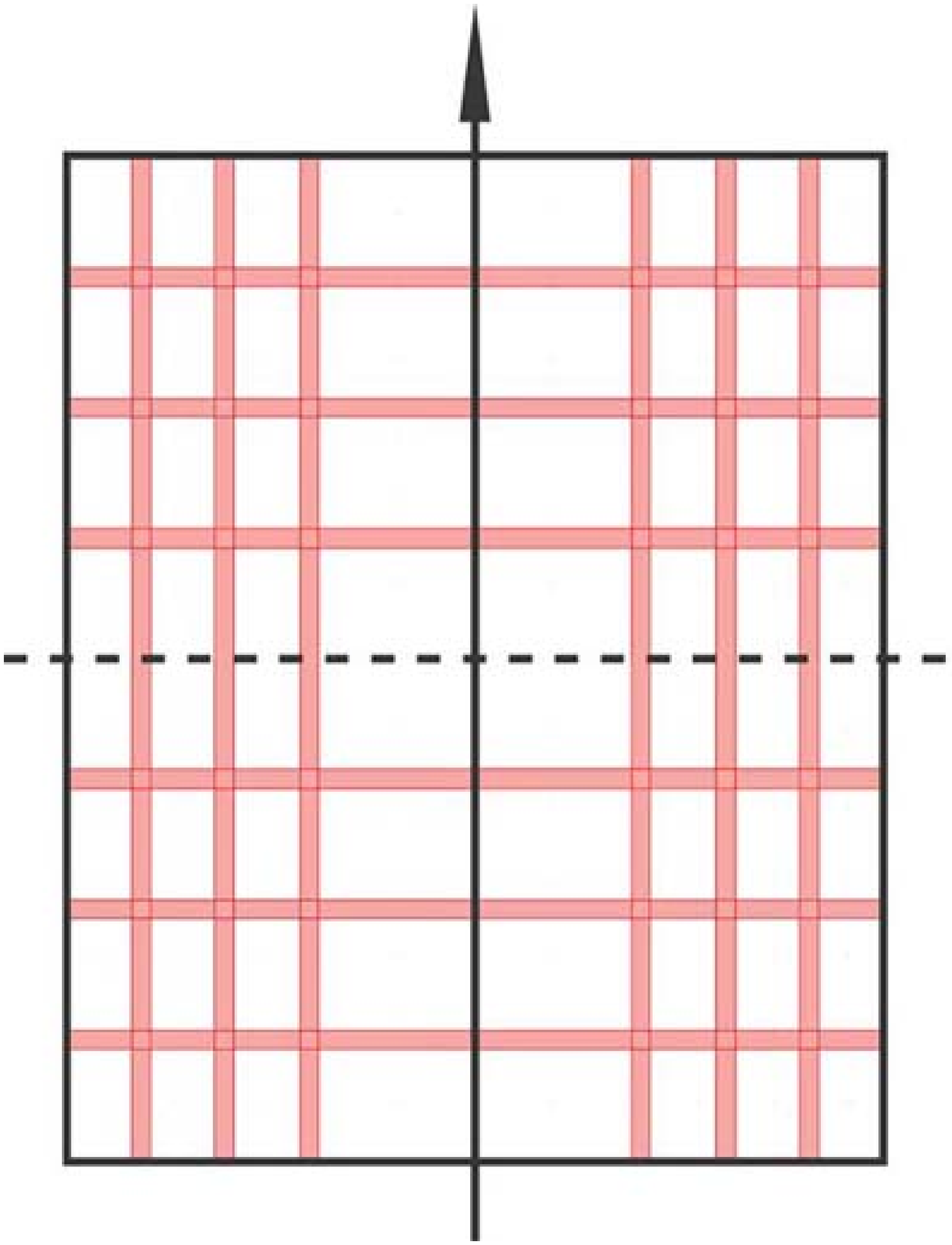}}&
b:$\:$\mbox{\includegraphics[width=0.42\columnwidth]{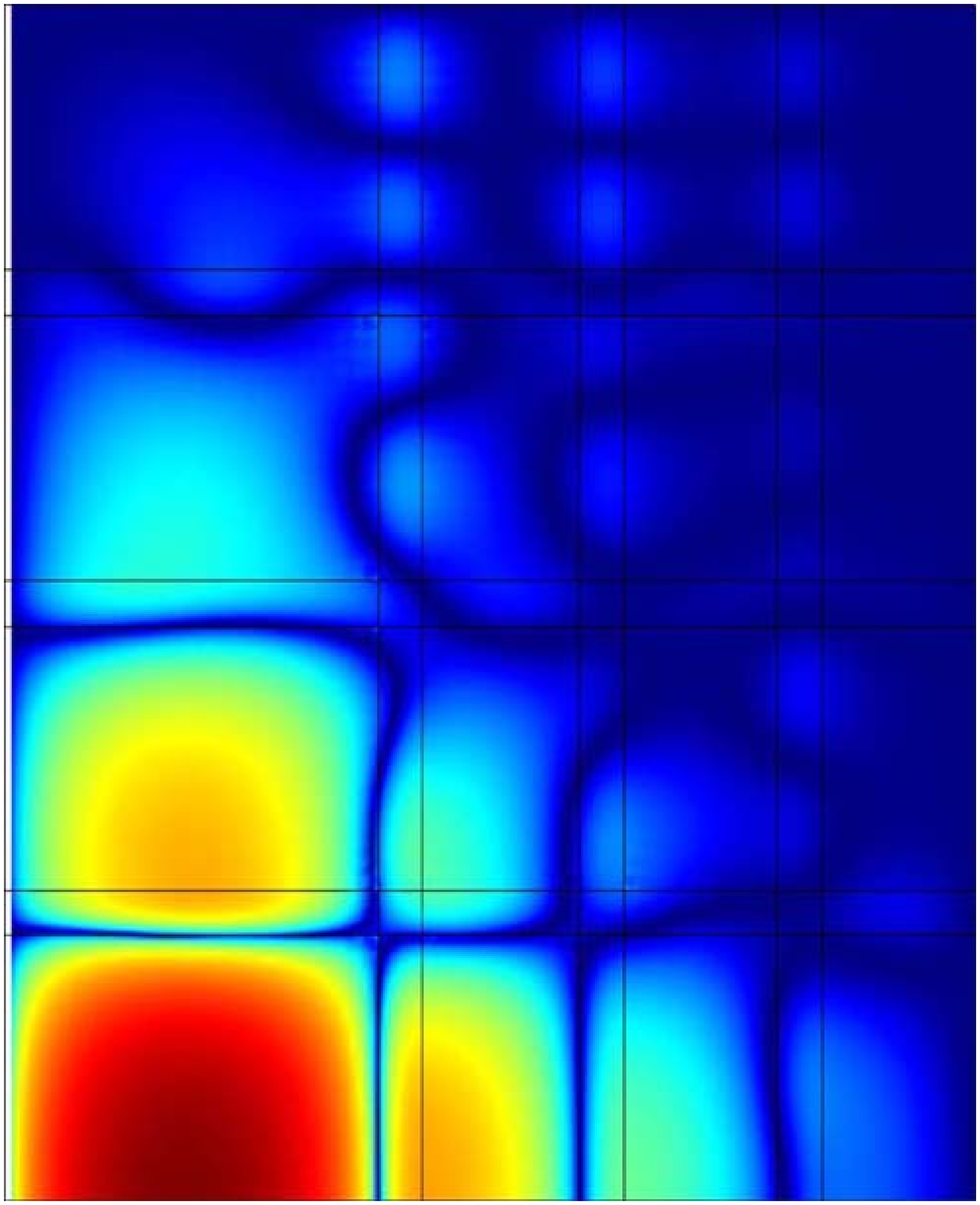}}
\end{tabular}
\caption{(a) Geometry (medial cross-section) of a 3rd-order distributed-Bragg-reflector
resonator within a cylindrical metallic can (hence electric interior walls
--represented by a solid black rectangle); as per ref.~\cite{flory97}, the can's
interior diameter is 10.98~cm and its interior height is 13.53~cm;
the horizontal and vertical grey (or pink --in color reproduction)
stripes denote cylindrical plates and barrels of sapphire;
white rectangles correspond to right cylinders/annuli of free-space;
the vertical arrow indicates the resonator's axis of rotational symmetry,
with which the sapphire crystal's c-axis is aligned;
a magnetic wall is imposed over the resonator's equatorial plane of
mirror symmetry (dashed horizontal line; cf.~M1 in Fig.~\ref{fig:generic_resonator}).
(b) False-color plot of the (logarithmic) electric-field intensity
$|\textbf{E}|^2$ over the top-right medial quadrant of the rotationally invariant
($M = 0$) TE$_{01}$ mode; note how the mode is strongly localized within the
resonator's central cylinder of free-space} \label{fig:BraggCavity}
\end{figure}
The resonator's model geometry was generated through an auxiliary script written in MATLAB.
Its corresponding mesh comprised 5476 base-mesh elements, with 66603 degrees
of freedom (DOF), with 8 edge vertices for each $\sim\!\!\lambda/4$ layer
of sapphire. Based on ref.~\cite{flory93}'s quartic fitting polynomials, the sapphire crystal's dielectric
permittivities, at a temperature $T = 300$~K, were set to $\epsilon_\perp  =  9.394$
(consistent with ref.~\cite{flory97}) and $\epsilon_\parallel = 11.593$.
The TE$_{01}$ mode shown in Fig.~\ref{fig:BraggCavity}(b) was found to lie at 10.00183~GHz,
in good agreement with F\&T's `precisely 10.00 GHz'. Using equation \ref{eq:filling_factor},
the mode's electric filling factor for the resonator's sapphire parts was 0.1270, which, assuming
an (isotropic) loss tangent of $5.9 \times 10^{-6}$ as per F\&T, corresponds to a
dielectric-loss $Q$ factor of 1.334 million.
Through equations \ref{eq:Q_wall_loss} and \ref{eq:length}, and assuming a surface resistance of 0.026 $\Omega$ per square
as per F\&T, the wall-loss $Q$ factor was determined to be 29.736 million, leading to a composite $Q$ (for dielectric and
wall losses operating in tandem) of 1.278 million. These three $Q$ values are consistent with F\&T's stated
(composite) $Q$ of `1.3 million, and limited entirely by dielectric losses'.
\begin{figure}[h]
\centering
\begin{tabular}{@{}l}
a:$\:$ \mbox{\includegraphics[width=0.50\columnwidth]{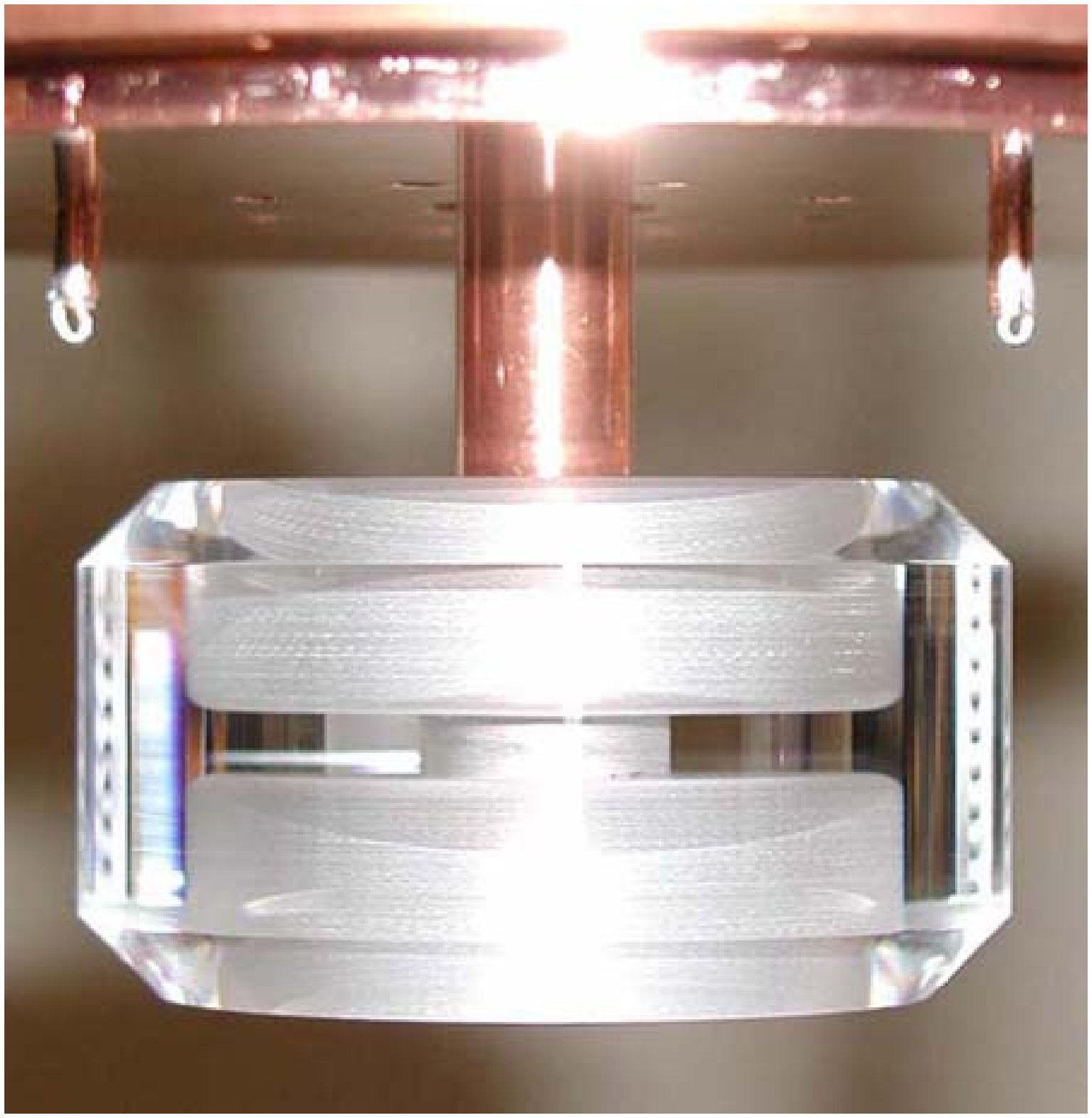}}\\
\\
\end{tabular}
\begin{tabular}{@{}l@{\quad}l}
b:$\:$ \mbox{\includegraphics[width=0.40\columnwidth]{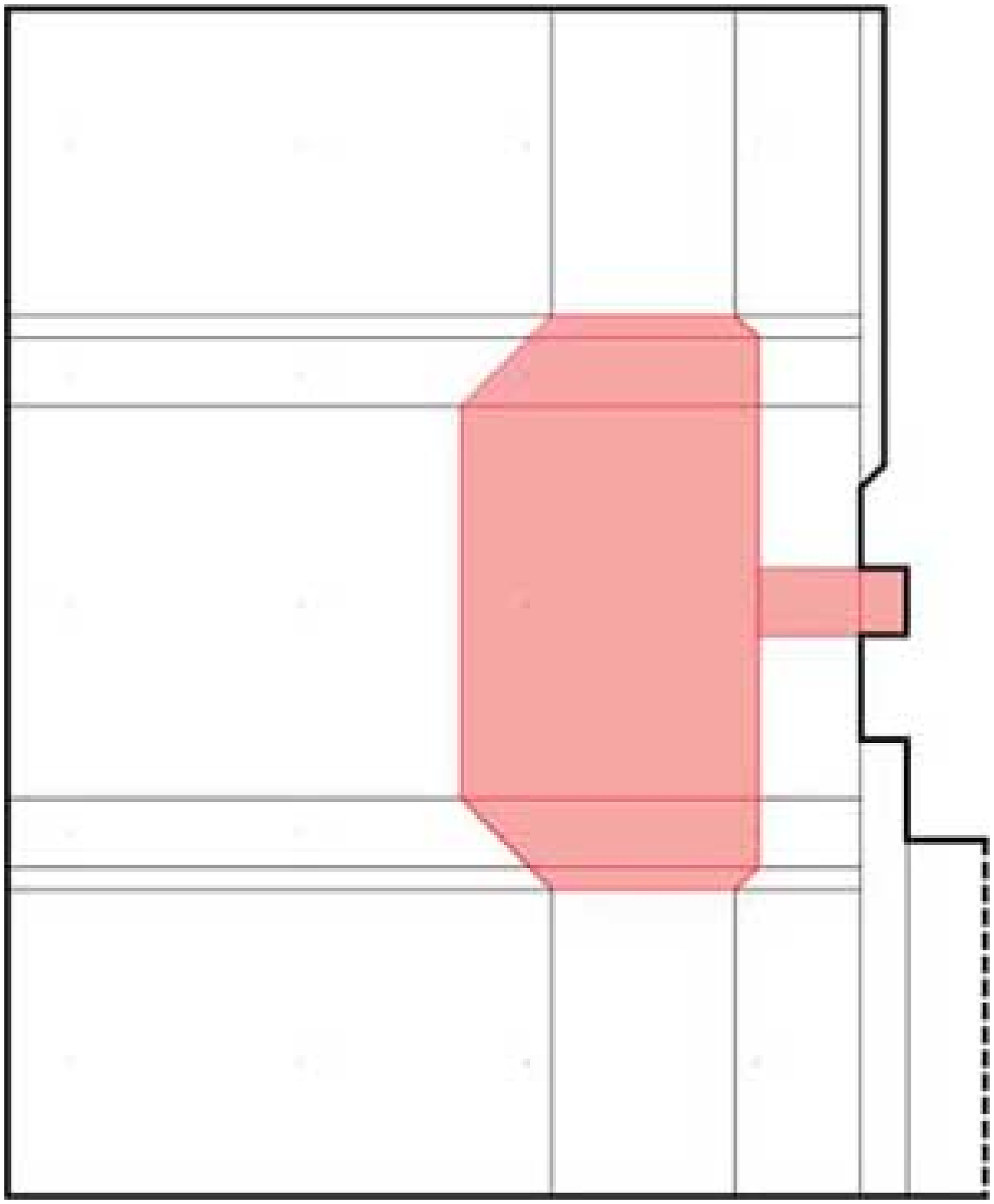}}
& c:$\:$ \mbox{\includegraphics[width=0.40\columnwidth]{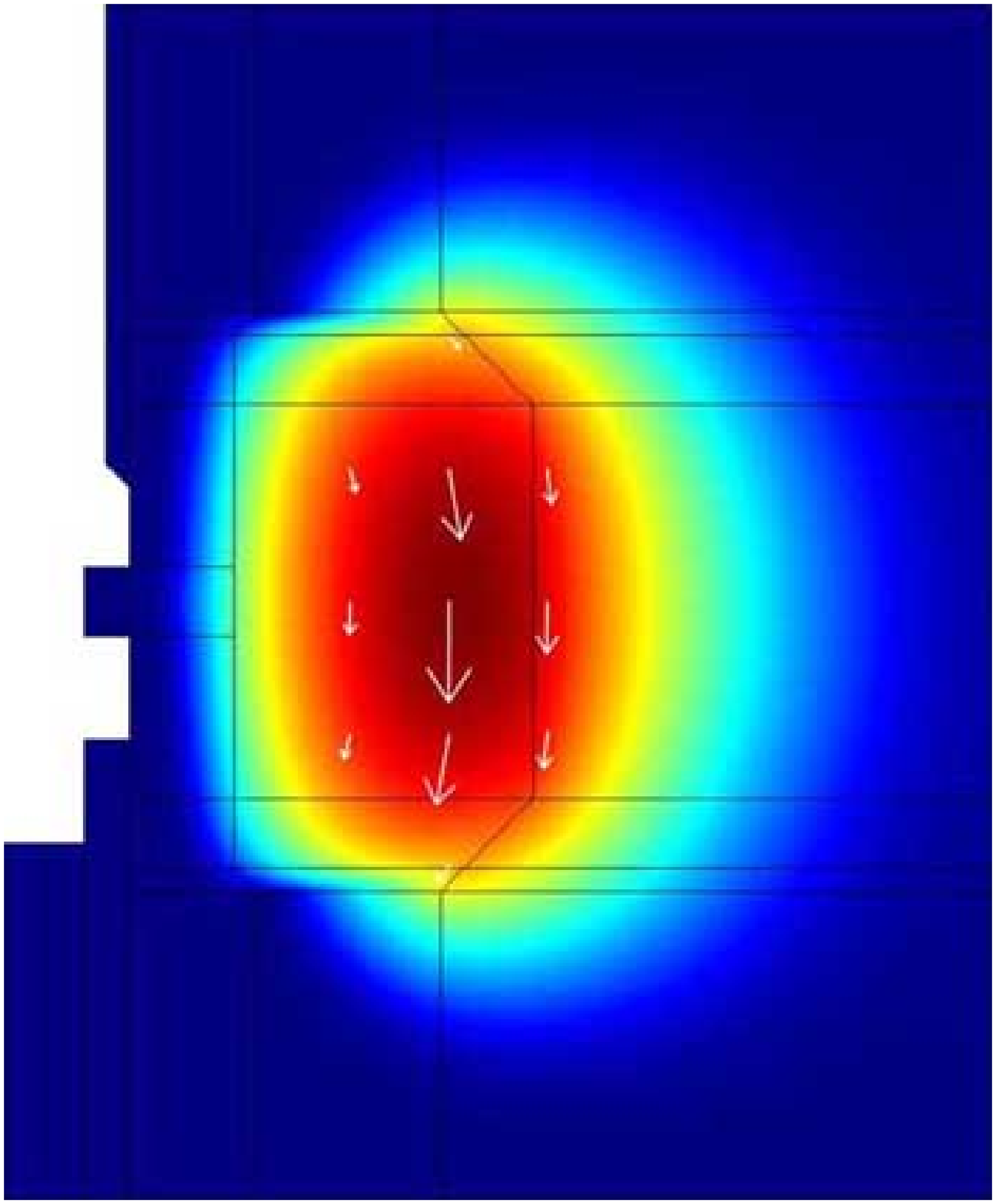}}\\
\end{tabular}
\caption{(a) Close-up of NPL's cryo-sapphire resonator, with the main body of its outer copper
can removed. The resonator's chamfered HEMEX sapphire ring has an outer diameter of $\sim$46.0~mm
and an axial height of $\sim$25.1~mm. The ring's integral interior `web', 3mm thick,
lies parallel to, and is centered (axially) on, the ring's equatorial plane, and
is supported through a central copper post.
Optical refraction at the ring's outer surface falsely exaggerates its internal diameter.
Above the ring are two loop probes for coupling electromagnetically to the resonator's
operational whispering-gallery mode.
(b) geometry of the resonator in medial cross-section; pink/grey indicates sapphire, white free space;
bounding these dielectric domains, and shown as solid black lines, are copper surfaces belonging to the
resonator's can and web-supporting post [the dashed vertical line (lower right) runs along the
resonator's cylindrical axis ($r = 0$)].
(c) false-color map (logarithmic scale) of $|\textbf{H}|^2$ for the resonator's 11th-azimuthal-mode-order
fundamental quasi-transverse-magnetic (N1$_{11}$ in
ref.~\cite{tobar01}'s notation) whispering-gallery mode at 9.146177 GHz (simulated),
as detailed on the 6th row of TABLE~\ref{tab:PermittivityFit}.
The white arrows indicate the magnitude and
direction of this mode's electric field strength ($\textbf{E}$) in the medial plane.}
\label{fig:NPLCsSapphRes}
\end{figure}
\section{Determination of the permittivities of cryogenic sapphire\label{sec:PermDet}}
\begin{table}[t]
\caption{\label{tab:PermittivityFit}NPL's cryogenic sapphire
resonator: simulated and experimental WG modes compared}
\begin{minipage}{\linewidth}
\renewcommand{\thefootnote}{thempfootnote}
\begin{tabular}{l@{\quad}l@{\quad}l@{\quad}l@{\quad}l@{\quad}l@{\quad}l@{\quad}l}
\hline
Simulated   &Simul.       &Simul.       &Mode      &Experi-     &Exper.  &Exper.   &Exper.\\
minus       &perp.     &para.       &ID\footnote{the nomenclature of ref.~\cite{tobar01} is used for this column.}
                                                    &mental    &width\footnote{full width half maximum (-3 dB).}
                                                                        &turn-    &Kram.\footnote{the difference
in frequency between the orthogonal pair of standing-wave resonances (akin to a
`Kramers doublet' in atomic physics) associated with the WG mode; the experimental parameters
stated in other columns correspond to the strongest resonance (greatest $S_{21}$ at line center)
of the pair.} \\
experim.    &filing      &filling      &          &freq.       &        &over   &split.\\
frequency   &factor      &factor       &          &            &        &temp.        &\\
$[$MHz$]$   &            &             &          &$[$GHz$]$   &$[$Hz$]$&$[$K$]$ &$[$Hz$]$\\
\hline
 -0.451     &0.860      &0.090      &S2$_6$      &6.954664    &285     &        &780\\
 -0.945     &0.930      &0.028      &S2$_7$      &7.696176    &82.5    &$<4.2$     &158\\
 0.881      &0.453      &0.517      &S4$_6$      &8.430800    &        &        &\\
 -1.538     &0.951      &0.014      &S2$_8$      &8.449908    &44.5    &$<4.2$     &418\\
 -0.412     &0.674      &0.299      &N2$_8$      &9.037458    &        &$4.8$     &\\
 -2.208     &0.071      &0.917      &N1$_{11}$   &9.148385    &9       &$5.0$     &57\\
 -1.916     &0.960      &0.009      &S2$_9$      &9.204722    &15.5    &$<4.2$     &88\\
 0.498      &0.251      &0.733      &S1$_{10}$   &9.267650    &12      &$5.2$     &180\\
 1.055      &0.287      &0.685      &N4$_8$      &9.421207    &80      &$5.0$     &\\
 -0.177     &0.437      &0.543      &S3$_8$      &9.800335    &84      &$4.8$     &1850\\
 0.358      &0.223      &0.763      &S1$_{11}$   &9.901866    &10      &$5.0$     &160\\
 -2.269     &0.965      &0.007      &S2$_{10}$   &9.957880    &24      &$<4.2$     &\\
 1.32       &0.730      &0.246      &S4$_8$      &10.27242    &153     &$5.0$     &\\
 0.19       &0.200      &0.787      &S1$_{12}$   &10.53863    &9.5     &$4.9$     &24\\
 0.00       &0.181      &0.808      &S1$_{13}$   &11.17728    &24.5    &$4.9$     &42\\
 4.13       &0.972      &0.006      &S2$_{12}$   &11.44918    &10      &$5.2$     &\\
\hline
\end{tabular}
\end{minipage}
\end{table}
\begin{figure}[t]
\centering
\mbox{\includegraphics[width=0.82\columnwidth]{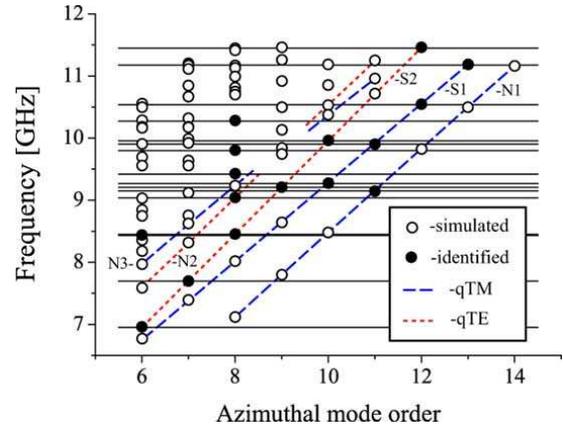}}
\caption{Plot used to aid the identification of experimental with simulated WG modes.
Solid horizontal lines (16 in total) indicate the center frequencies of the former. Solid circles
indicate the identification of a simulated mode with an experimental one (the difference
in their frequencies corresponds to much less than a circle's radius in all cases);
hollow circles indicate simulated modes that were not identified with any experimentally
measured one. Quasi-transverse-magnetic (q-TM) and quasi-transverse-electric (q-TE) WG modes
of the same family are joined by (blue-)dashed and (red-)dotted lines respectively; a few of the
lowest-lying mode families are labeled using standard notation \cite{tobar01}.}
\label{fig:ExpvSimCsRes}
\end{figure}
The method is here applied to determine the two dielectric constants of monocrystalline
sapphire (HEMEX grade \cite{crystalsystems06}) at 4.2~K from
a set of experimental data,
listed in the four right-most columns of
TABLE~\ref{tab:PermittivityFit},
and corresponding to the resonator whose innards
are shown in Fig.~\ref{fig:NPLCsSapphRes}(a).
Allowance was made for the shrinkages of the resonator's constituent materials
from room to liquid-helium temperature\footnote{By integrating
up published linear-thermal-expansion data
(\emph{viz.} Table~4/TABLE~1 of ref.~\cite{white83}/\cite{white93}),
sapphire's two cryo-shrinkages were estimated to be
$(1.0 - 7.21 \times 10^{-4})$ and $(1.0 - 5.99 \times 10^{-4})$ for directions
parallel and perpendicular to the sapphire's c-axis, respectively.
From Table~F at the back of ref.~\cite{white79}, the cryo-shrinkage
of (isotropic) copper was taken to be $(1.0 - 3.26 \times 10^{-3})$.}
and the values of sapphire's two dielectric constants
($\epsilon_{\perp}$ and $\epsilon_{\parallel}$)
were initially set equal to those specified in ref.~\cite{krupka99a}.
Fig.~\ref{fig:NPLCsSapphRes}(b)'s geometry was meshed with quadrilaterals over
the medial half-plane, with 8944 elements in its base mesh, and with DOF~$= 108555$.
For a given azimuthal mode order $M$, calculating the lowest 16 eigenmodes took
around 3 minutes on the author's office PC (as previously specified).
With Fig.~\ref{fig:ExpvSimCsRes} as a guide,
each of the 16 experimental resonances was identified to a particular simulated WG mode,
as specified in the 4th column of TABLE~\ref{tab:PermittivityFit}, lying near to it in frequency;
these identifications were influenced by requiring that the measured attributes (\emph{e.g.}
the FWHM linewidths) of the experimental resonances belonging --as per their identifications-- to the
same `family' of WG modes (\emph{e.g.}~S1, or N2) varied smoothly with $M$.
Filling factors were then calculated to quantify the differential change in the frequency
of each identified mode with respect to $\epsilon_{\perp}$ and $\epsilon_{\parallel}$.
The two dielectric constants were then
adjusted to minimize the $\chi^2$ variance of the residual (simulated-minus-measured) frequency
differences. The resultant best-fit values, to which the simulated data
occupying the three left-most columns in TABLE~\ref{tab:PermittivityFit} correspond, were:
\begin{eqnarray}
\label{eq:fitted_e_perp}
\epsilon_\perp & = & 9.285 \,\,(\pm 0.010);\\
\label{eq:fitted_e_para}
\epsilon_\parallel  & = &11.366 \,\,(\pm 0.010).
\end{eqnarray}
The nominal error assigned to each reflects uncertainties in the identifications
of certain experimental resonances, each lying almost equally close
(in both frequency and other attributes) to two or more different
simulated WG modes.
Errors resulting from a finite meshing density~\cite{santiago94},
or those from the finite dimensional/geometric tolerances to which
the resonator's shape was known, were estimated to be small in comparison.

\section*{Acknowledgments}
The author thanks Anthony Laporte and Dominique Cros at XLIM, Limoges,
France, for
an independent (and corroborating) 2D-FEM
simulation of the resonator considered in section~\ref{sec:PermDet},
and Jonathan Breeze at Imperial College, London,
for suggesting the DBR resonator analyzed in subsection~\ref{subsec:BraggCav}.
He also thanks three NPL colleagues:
Giuseppe Marra, for
some of the experimental data presented in Table~\ref{tab:PermittivityFit},
Conway Langham for his
estimated values for the cryo-shrinkages of sapphire,
and Louise Wright, for a detailed review of an early manuscript.
%




\bstctlcite{IEEE:BSTcontrol}

\bibliographystyle{IEEEtran}
%
%
%

%

\begin{biography}[{\includegraphics[width=1in,height=1.25in,clip,keepaspectratio]{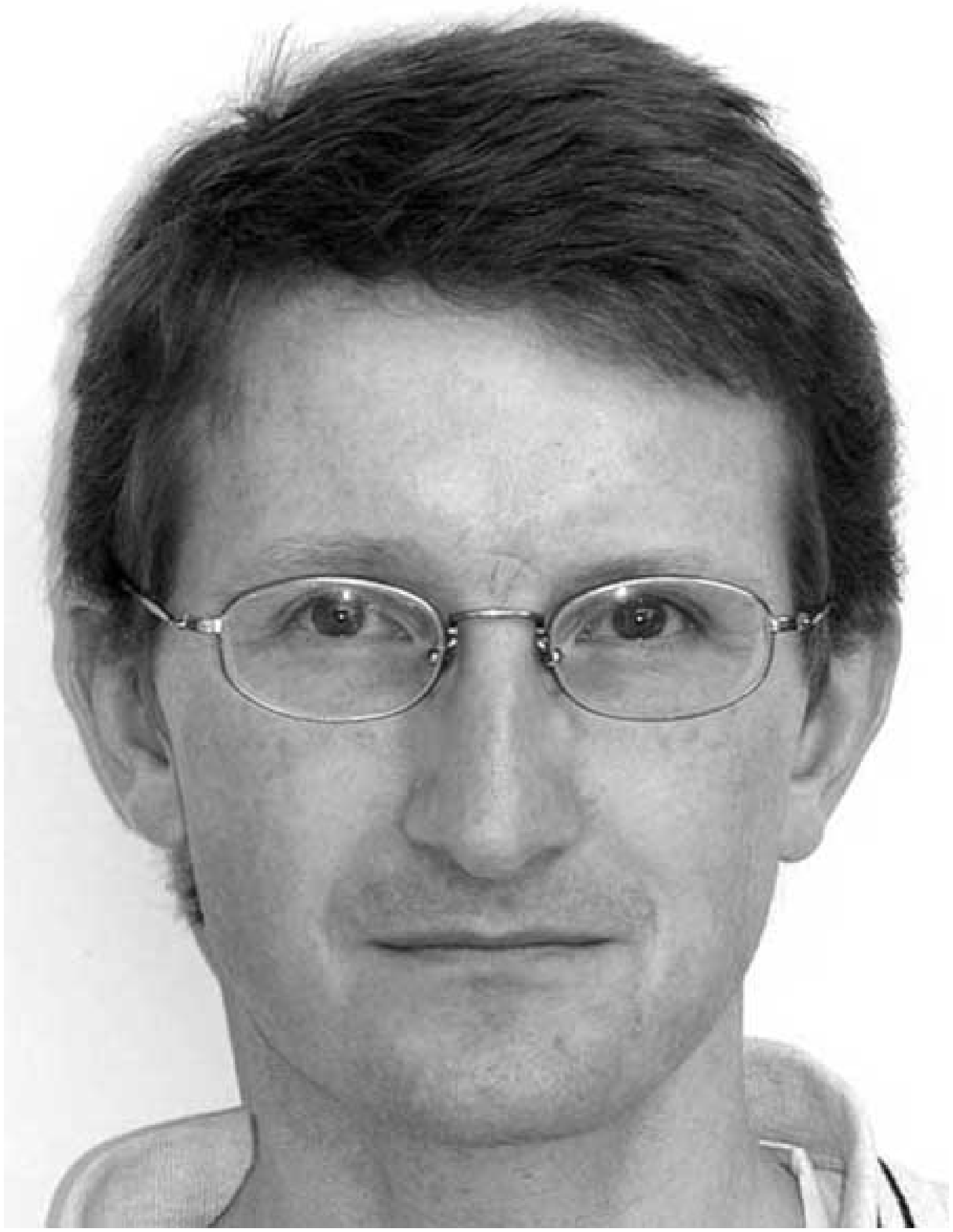}}]{Mark Oxborrow}
was born near Salisbury, England, in 1967. He received a B.A. in physics from the University of Oxford in 1988, and
a Ph.D. in theoretical condensed-matter physics from Cornell University in 1993; his thesis topic concerned
random-tiling models of quasicrystals.
During subsequent postdoctoral appointments at both the Niels Bohr Institute in Copenhagen and then back
at Oxford, he investigated acoustic analogues of quantum wave-chaos.
In 1998, he joined the UK's National Physical Laboratory; his eclectic, project-based research work there to
date has included the design and construction of ultra-frequency-stable microwave and optical oscillators,
the development of single-photon sources, and the applications of carbon nanotubes to metrology.
\end{biography}


\end{document}